\documentclass[11pt]{article}
\usepackage{}
\usepackage{amssymb}
\usepackage{amsfonts}
\usepackage{mathrsfs}
\usepackage{graphicx}
\usepackage{amsmath}
\usepackage{color}
\usepackage{amsthm}
\usepackage{epstopdf}
\usepackage{pifont,bm}
\usepackage{tikz}
\usepackage{enumerate}
\theoremstyle{definition}

\makeatletter
\def\@biblabel#1{[#1]}
\makeatother

\makeatletter \@addtoreset{equation}{section}

\begin{document}
%\begin{CJK*}{GBK}{song}

\begin{titlepage}
\title{\bf{On the inverse scattering transform to the discrete Hirota equation with nonzero boundary conditions
\footnote{This work is supported
by the National Natural Science Foundation of China (No.12271129 and No.12201622) and the China Scholarship Council (No.202206120152).\protect\\
\hspace*{3ex}$^{*}$Corresponding authors.\protect\\
\hspace*{3ex} E-mail addresses: guixianwang@hit.edu.cn (G.X. Wang), xbwang@cumt.edu.cn (X.-B. Wang), bohan@hit.edu.cn (B. Han)}
}}
\author{Guixian Wang$^{1}$, Xiu-Bin Wang$^{2}$, Bo Han$^{1,*}$\\
%%%%%%%%%%%%%%%%%%%%%%%%%%%%%%%%%%%%%%%%%%%%%%%%%%%%%%%%%%%%%%%%%%%%%%%%%%%%%%%%%%%%%%%%%
%%%%%              以下两行为作者单位
%%%%%%%%%%%%%%%%%%%%%%%%%%%%%%%%%%%%%%%%%%%%%%%%%%%%%%%%%%%%%%%%%%%%%%%%%%%%%%%%%%%%%%%%%
\small \emph{$^{1}$Department of Mathematics, Harbin Institute of Technology, Harbin 150001, China} \\
%\small \emph{China}\\
\small \emph{$^{2}$School of Mathematics and Institute of Mathematical Physics,}\\
\small \emph{China University of Mining and Technology, Xuzhou 221116, China} \\
\date{}}
\thispagestyle{empty}
\end{titlepage}
\maketitle

\vspace{-0.5cm}
\begin{center}
\rule{15cm}{1pt}\vspace{0.3cm}

\parbox{15cm}{\small
{\bf Abstract}\\
\hspace{0.5cm}
Under investigation in this work is the robust inverse scattering transform of the discrete Hirota equation with nonzero boundary conditions,
which is applied to solve simultaneously arbitrary-order poles on the branch points and spectral singularities.
Using the inverse scattering transform method, we construct the Darboux transformation but not with the limit progress, which is more convenient than before.
Several kinds of rational solutions are derived in detail. These solutions contain W-shape solitons, breathers, high-order rogue waves, and various interactions between solitons and breathers. Moreover, we analyze some remarkable characteristics of rational
solutions through graphics. %In particular, as parameters $a$ and $b$ change, the propagation angles and shapes of waves have changed observably.
Our results are useful to explain the related nonlinear wave phenomena.
}

\vspace{0.5cm}
\parbox{15cm}{\small{

\vspace{0.3cm} \emph{Key words:} Discrete Hirota equation, Inverse scattering transform, Riemann-Hilbert problem, Darboux transformation

 } }
\end{center}
\vspace{0.3cm} \rule{15cm}{1pt} \vspace{0.2cm}

\section{Introduction}

The Hirota equation reads
\begin{align}\label{HE}
\textrm{i}u_{\tau}+\alpha(u_{xx}\pm2|u|^2u)+\textrm{i}\beta(u_{xxx}\pm6|u|^2u_{x})=0,
\end{align}
which was originally introduced by Hirota \cite{R-1973}, where $\alpha$ and $\beta$ are real constants, and stand for the second-order and third-order dispersions, respectively.
%and has been studied by many researchers \cite{M-1983,N-1991,VI-2001,A-2010}.
Eq. \eqref{HE} is completely integrable, and is used to characterize a number of physical problems, such as vortex motion \cite{GL-1980}, the one-dimensional
Heisenberg spin system \cite{DG-1989}, and nonlinear optics \cite{LF-1980}. Using the Ablowitz-Ladik's
formulation \cite{MJ-1976}, Porsezian and Lakshmanan \cite{K-1989} proposed the discrete Hirota equation with form
\begin{align}\label{sdHE}
\textrm{i}q_{n,t}=(1+|q_{n}|^{2})\left[(a+\textrm{i}b)q_{n+1}+(a-\textrm{i}b)q_{n-1}\right]-2aq_{n},
\end{align}
which is a generalized nonlinear Schr\"{o}dinger equation (NLSE) with high-order terms, in which $a$ and $b$ are constant parameters.
%System \eqref{sdHE} is a combination of discrete NLSE and discrete complex modified KdV equation (cmKdV), in other words, when $b = 0$, Eq.\eqref{sdHE} converts into the discrete NLSE, and it turns into discrete cmKdV equation by taking $a = 0$.

The inverse scattering transform (IST) was first presented by Gardner et al. and was used to solve KdV equation with initial-valued problems \cite{GS-1967}. There are many works about integrable systems with initial-valued problems by means of IST. Zakharov and Shabat studied NLSE via IST in 1972 \cite{VE-1972}.
In 1974, Ablowitz et al. investigated the Ablowitz Kaup Newell Segur system, and after that a great number of integrable systems have been solved by IST \cite{MJ-1973,MJ-1974,MJ-1975}. IST method is originally based on the classical Gelfand-Levitan-Marchenko (GLM) equation, and the GLM equation is replaced by the Riemann-Hilbert problem (RHP) \cite{P-1993,A-2003,XB-2020-1,XB-2020}, which greatly simplifies the solution process. IST also has been applied to discrete integrable systems with nonzero boundary conditions (NZBCs) \cite{VE-1992,OA-1992,MJ-2007,HQ-2018,AK-2019,MS-2021,Y-2021,GX-2023}. Recently, Bilman and Miller \cite{DB-2019,DB-2019-1,DB-2020} have presented a robust IST to solve the rogue wave solutions, which aims to regain the meromorphic matrix function in terms of the normalization approach.
%Later, as a modern version of IST, the RHP for integrable
%systems was presented.27 It has become clear that the RHP is applicable to
%the construction of exact solutions and asymptotic analysis of solutions for a wide class of integrable systems.28-35

Eq. \eqref{sdHE} is a completely integrable model, and
some properties have attracted wide attention. In \cite{K-1989}, Porsezian
and Lakshmanan constructed the B\"{a}cklund transformation (BT) and obtained the one-soliton solutions on the basis of IST. In \cite{R-2016}, new discrete one-soliton solutions were derived via the obtained discrete Darboux transformation (DT). Zhu et al. showed that the gauge equivalent equations of the nonintegrable discrete Hirota equations under the discrete gauge transformations and studied the dynamical properties \cite{LY-2014}. A new discrete Hirota equation which can generate the Hirota equation in the continuum limit was presented in \cite{A-2016}, authors deduced first order, second order exact soliton solutions in view of constructing the DT.

Since the traditional scattering analysis for the rogue wave solution is based the same scattering data with boundary conditions, it is not suitable for the study of high-order rogue waves. It is natural to adopt the robust IST method to calculate the general high-order rogue wave solutions. Furthermore, we construct the DT under the robust IST and obtain these solutions without taking the limit on the  spectral parameter.
There has been no studies of high-order rogue waves for Eq. \eqref{sdHE} with NZBCs. Considering the modulational instability effect \cite{TB-1967}, we introduce the parameter $B$ into the expression of NZBCs
\begin{align}\label{AF-1}
\lim_{n\rightarrow\pm\infty} q_{n}=q_{n}^{\pm}\equiv A\textrm{e}^{\textrm{i}[B(n+\frac{1}{2})+C t+B_{\pm}]},
\end{align}
with
\begin{align}\nonumber
C=2a[1-(1+A^{2})\cos B]+2b(1+A^{2})\sin B,
\end{align}
where parameters $A$, $B$ and $B_{\pm}$ are real constants.
When $B=\frac{\pi}{2}+n\pi$ $(n\in \mathbb{Z})$, the NZBCs \eqref{AF-1} are modulational stable, there exist periodic solutions or W-shape soliton solutions. When $B\neq\frac{\pi}{2}+n\pi$ $(n\in \mathbb{Z})$, the NZBCs \eqref{AF-1} are modulational unstable, there exist a series of breather waves and rogue waves.

The outline of the paper is as follows. Sec. 2 aims to investigate the discrete Hirota equation \eqref{sdHE} with the NZBCs \eqref{AF-1} via the robust IST. Firstly, we need to find a new RHP to obtain the solutions with spectral singularity. Then we normalize the RHP with a diagonal matrix which is only related to $n$ and $t$ in the neighborhood of $\infty$ and $0$. Considering the loop group approach, the fundamental DT is constructed in Sec. 3. We obtain the general Darboux matrix with the forms of the RHP by using the robust IST. Moreover, we conclude the compact rational solutions.
%Sec.3 devotes to construct the elementary Darboux matrix in terms of the loop group method. We obtain the general Darboux matrix with the forms of the RHP by using the robust IST. Moreover, we conclude the compact rational solutions.
In Sec. 4, we deduce the expressions of W-shape solitons, breathers as well as high-order rogue waves and calculate the highest crest value of rational solutions by means of the BT. Then dynamic behaviors of these solutions are shown visually through figures. Finally, we give some conclusions and discussions.

\section{The robust inverse scattering transform}

Eq. \eqref{sdHE} admits the following Lax pair
\begin{align}\label{AF-2}
\phi_{n+1}=U_{n}\phi_{n},~~\phi_{n,t}=V_{n}\phi_{n},
\end{align}
with
\begin{small}
\begin{equation}\label{AF-2-1}
\begin{split}
&U_{n}=\left(
         \begin{array}{cc}
           \lambda & q_{n} \\
           -\bar{q}_{n} & \frac{1}{\lambda} \\
         \end{array}
       \right)
,~~\lambda\in \mathbb{C},\\
&V_{n}=\left(
         \begin{array}{cc}
           (b-\textrm{i}a)q_{n}\bar{q}_{n-1}-\frac{\textrm{i}a}{2}(\lambda-\lambda^{-1})^{2}+b\lambda^{2} & (b-\textrm{i}a)q_{n}\lambda+(b+\textrm{i}a)q_{n-1}\lambda^{-1} \\
           -(b+\textrm{i}a)\bar{q}_{n}\lambda^{-1}-(b-\textrm{i}a)\bar{q}_{n-1}\lambda & (b+\textrm{i}a)\bar{q}_{n}q_{n-1}+\frac{\textrm{i}a}{2}(\lambda-\lambda^{-1})^{2}+b\lambda^{-2} \\
         \end{array}
       \right),
\end{split}
\end{equation}
\end{small}
where the spectral parameter $\lambda$ belongs to $\mathbb{C}$, $q_{n}$ denotes the potential function, and the complex conjugation is expressed by superscript $``-"$.

It is straightforward to check that the expressions \eqref{AF-2-1} satisfy the compatibility condition
\begin{align}\nonumber
U_{n,t} + U_{n}V_{n}-V_{n+1}U_{n} = 0.
\end{align}
Under the following transformation
\begin{align}\nonumber
q_{n} =v_{n}\textrm{e}^{\textrm{i}[B(n+\frac{1}{2})+C t]},~~\psi_{n}=\textrm{e}^{-\frac{\textrm{i}}{2}[B n+C t]\sigma_{3}}\phi_{n},~~\lambda=z\textrm{e}^{\frac{\textrm{i}B}{2}},
\end{align}
with $\sigma_{3}=\mathrm{diag}\left(1,-1\right)$,
%\begin{align}\nonumber
%\sigma_{3}=\left(
%             \begin{array}{cc}
%               1 & 0 \\
%               0 & -1 \\
%             \end{array}
%           \right).
%\end{align}
Eq. \eqref{sdHE} turns into
\begin{equation}\label{AF-3}
\begin{split}
\textrm{i}v_{n,t}=&(1+|v_{n}|^{2})\left[(a+\textrm{i}b)v_{n+1}\textrm{e}^{\textrm{i}B}
+(a-\textrm{i}b)v_{n-1}\textrm{e}^{-\textrm{i}B}\right]\\
&-2(1+A^{2})(a\cos B-b\sin B)v_{n},
\end{split}
\end{equation}
and the NZBCs \eqref{AF-1} transform into
\begin{align}\label{AF-5}
\lim_{n\rightarrow\pm\infty} v_{n}=v_{n}^{\pm}\equiv A\textrm{e}^{\textrm{i}B_{\pm}}.
\end{align}
We rewrite the Lax pair \eqref{AF-2} as
\begin{align}\label{AF-4}
\psi_{n+1}=X_{n}\psi_{n},~~\psi_{n,t}=T_{n}\psi_{n},
\end{align}
with
\begin{equation}\nonumber
\begin{aligned}
&X_{n}=\left(
         \begin{array}{cc}
           z & v_{n} \\
           -\bar{v}_{n} & z^{-1} \\
         \end{array}
       \right)\notag\\
      &~~~~=z\left(
                        \begin{array}{cc}
                          1 & 0 \\
                          0 & 0 \\
                        \end{array}
                      \right)+\left(
         \begin{array}{cc}
           0 & v_{n} \\
           -\bar{v}_{n} & 0 \\
         \end{array}
       \right)+\left(
                        \begin{array}{cc}
                          0 & 0 \\
                          0 & 1 \\
                        \end{array}
                      \right)z^{-1}\notag\\
      &~~~~=z M_{+}+N_{n}+M_{-}z^{-1},\notag\\
      &T_{n}=\left(
         \begin{array}{cc}
           T_{n}^{(11)} & T_{n}^{(12)} \\
           T_{n}^{(21)} & T_{n}^{(22)} \\
         \end{array}
       \right),
\end{aligned}
\end{equation}
where
\begin{equation}\nonumber
\begin{aligned}
&T_{n}^{(11)}:=(b-\textrm{i}a)v_{n}\bar{v}_{n-1}\textrm{e}^{\textrm{i}B}-\frac{\textrm{i}a}{2}(z \textrm{e}^{\frac{\textrm{i}B}{2}}-z^{-1}\textrm{e}^{\frac{-\textrm{i}B}{2}})^{2}+bz^{2} \textrm{e}^{\textrm{i}B}-\frac{\textrm{i}C}{2},\notag\\
&T_{n}^{(12)}:=(b-\textrm{i}a)v_{n}z\textrm{e}^{\textrm{i}B}+(b+\textrm{i}a)v_{n-1}z^{-1} \textrm{e}^{-\textrm{i}B},\notag\\
&T_{n}^{(21)}:=-(b+\textrm{i}a)\bar{v}_{n}\textrm{e}^{-\textrm{i}B}z^{-1}-(b-\textrm{i}a)
\bar{v}_{n-1}\textrm{e}^{\textrm{i}B}z,\notag\\
&T_{n}^{(22)}:=(b+\textrm{i}a)\bar{v}_{n}v_{n-1}\textrm{e}^{-\textrm{i}B}+\frac{\textrm{i}a}{2}(z \textrm{e}^{\frac{\textrm{i}B}{2}}-z^{-1} \textrm{e}^{\frac{-\textrm{i}B}{2}})^{2}+bz^{-2}\textrm{e}^{-\textrm{i}B}+\frac{\textrm{i}C}{2}.
\end{aligned}
\end{equation}
%Therefore, Eq.\eqref{AF-3} possesses nonzero boundary conditions with the form
%Then the Cauchy problem for \eqref{sdHE} corresponds to the Cauchy problem for \eqref{AF-3} with boundary condition
%\begin{align}\label{AF-5}
%\lim_{n\rightarrow\pm\infty} v_{n}=v_{n}^{\pm}\equiv A\textrm{e}^{\textrm{i}B_{\pm}}.
%\end{align}

\subsection{Scattering problems}
We only need to analyze the spectral problem \eqref{AF-4} for direct scattering problem. Firstly, we have
\begin{align}\label{AF-6}
\psi_{n}^{\pm}=\textrm{e}^{\frac{\textrm{i}}{2}B_{\pm}\sigma_{3}}\chi_{n}^{\pm}.
\end{align}
It follows from \eqref{AF-4} and \eqref{AF-6} that we calculate
\begin{align}\label{AF-7}
\chi_{n+1}^{\pm}=(Z+N_{n}^{\pm})\chi_{n}^{\pm},
\end{align}
with
\begin{align}\nonumber
Z=\left(
          \begin{array}{cc}
            z & A \\
            -A & z^{-1} \\
          \end{array}
        \right),
~~N_{n}^{\pm}=\left(
                \begin{array}{cc}
                  0 & v_{n}\textrm{e}^{-\textrm{i}B_{\pm}}-A \\
                  -\bar{v}_{n}\textrm{e}^{\textrm{i}B_{\pm}}+A & 0 \\
                \end{array}
              \right).
\end{align}
Then we diagonalize the matrix $Z$ with the form
\begin{align}\nonumber
Z=rT\zeta^{\sigma_{3}}T^{-1},
\end{align}
where
\begin{equation}\nonumber
\begin{aligned}
&T=\left(
     \begin{array}{cc}
       1 & \xi \\
       \xi & 1 \\
     \end{array}
   \right),
~~\zeta^{\sigma_{3}}=\left(
                                                     \begin{array}{cc}
                                                       \zeta & 0 \\
                                                       0 & \zeta^{-1} \\
                                                     \end{array}
                                                   \right),\notag\\
&r=(1+A^{2})^{\frac{1}{2}},~~\xi=\frac{1-z^{2}+\sqrt{(1-z^{2})^{2}
-4A^{2}z^{2}}}{2Az}.
\end{aligned}
\end{equation}
Since $\zeta$ obeys
\begin{align}\nonumber
r(\zeta+\zeta^{-1})=z+z^{-1},
\end{align}
we have
\begin{align}\nonumber
\zeta=\frac{1+z^{2}+\sqrt{(1+z^{2})^{2}
-4r^{2}z^{2}}}{2rz}.
\end{align}
It follows from equations $\zeta^{2}=1$ and $\xi^{2}=1$ that we arrive at $z=r\pm A$ and $z=-r\pm A$.
%\begin{align}\nonumber
%\zeta^{2}=1,~~\xi^{2}=1,
%\end{align}
%we calculate
%\begin{align}\nonumber
%z=r\pm A,~~z=-r\pm A.
%\end{align}
The Riemann surface $M$ which possesses genus 1 is given by $\zeta(z)$. It divides into two pieces $M_{+}$ and $M_{-}$, along
\begin{align}\nonumber
\Sigma= [-(r + A), -(r-A)] \cup [(r-A), (r + A)].
\end{align}
The jump condition of $\zeta(z)$ is
\begin{align}\nonumber
\zeta_{-}=\zeta_{+}\zeta_{-}^{2},~~\zeta_{\pm}=\zeta(z\pm \textrm{i}0^{+}),
\end{align}
and so is $\xi(z)$.
\vspace{0.2cm}

\begin{figure}
\begin{center}
\centerline{\begin{tikzpicture}
\tikzstyle{arrow} = [->,>=stealth]
\filldraw[pink, line width=0](4.5,0) circle [radius=2];
\draw[line width=1pt](4.5,0) circle [radius=2];
\draw[thick](-7,0)--(0,0);
\draw[thick](1,0)--(8,0);
\draw[thick](-3.5,-3)--(-3.5,3);
\draw[thick](4.5,-3)--(4.5,3);
\filldraw[white, line width=0](4.5,0) circle [radius=0.5];
\draw[line width=1pt](4.5,0) circle [radius=0.5];
\draw[fill] (5.3,0.7) node[above]{$M$};
\draw[fill] (6.5,1.5) node[above]{$M_{+}$};
\draw[fill] (4.5,-0.3) node[above]{$M_{-}$};
\draw [-][arrow][dashed,thick](3.8,-1.3)--(3.2,-0.1);
\draw [-][arrow][dashed,thick](3.8,-1.3)--(5.8,-0.1);
\draw[fill] (3.8,-1.3) node[below]{$\Sigma$};
\draw[fill] (3.2,0) node[above]{$+$};
\draw[fill] (3.2,0) node[below]{$-$};
\draw[fill] (5.8,0) node[above]{$+$};
\draw[fill] (5.8,0) node[below]{$-$};
\draw[fill] (-3.5,-0.2) node[left]{$o$};
\draw[arrow][thick](-0.05,0)--(-0,0);
\draw[arrow][thick](-3.5,2.95)--(-3.5,3);
\draw[arrow][thick](7.95,0)--(8,0);
\draw[arrow][thick](4.5,2.95)--(4.5,3);
\draw[line width=1pt](-3.5,0) circle [radius=2];
\draw[fill] (-0.2,0) node[below]{$\mathrm{Re}(z)$};
\draw[fill] (-3.5,2.8) node[right]{$\mathrm{Im}(z)$};
\draw[fill] (7.7,0) node[below]{$\mathrm{Re}(z)$};
\draw[fill] (4.5,2.8) node[right]{$\mathrm{Im}(z)$};
\draw[fill] (-6.4,0) node[above]{$-(r+A)$};
\draw[fill] (-4.8,0) node[above]{$-r+A$};
\draw[fill] (-2,0) node[above]{$r-A$};
\draw[fill] (-0.6,0) node[above]{$r+A$};
\draw [-][arrow][dashed,thick](-5.8,-1)--(-5.8,-0.1);
\draw[fill] (-5.8,-1) node[below]{$\Sigma_{+}$};
\draw [-][arrow][dashed,thick](-1.2,-1)--(-1.2,-0.1);
\draw[fill] (-1.2,-1) node[below]{$\Sigma_{+}$};
\draw [-][arrow][dashed,thick](-4.2,-1)--(-5.3,-0.1);
\draw [-][arrow][dashed,thick](-4.2,-1)--(-1.7,-0.1);
\draw[fill] (-4.2,-1) node[below]{$\Sigma_{-}$};
\draw[fill][thick][color=blue][line width=2pt](-6.3,0) -- (-4.8,0);
\draw[fill][thick][color=blue][line width=2pt](-2.2,0) -- (-0.7,0);
\draw[fill][thick][color=blue][line width=2pt](6.2,0) -- (5.4,0);
\draw[fill][thick][color=blue][line width=2pt](2.8,0) -- (3.6,0);
\draw[arrow][thick][color=blue][line width=2pt](2.8,0) -- (3.4,0);
\draw[arrow][thick][color=blue][line width=2pt](5.4,0) -- (6,0);
\draw[fill] (-6.3,0) circle [radius=0.05];
\draw[fill] (-4.8,0) circle [radius=0.05];
\draw[fill] (-2.2,0) circle [radius=0.05];
\draw[fill] (-0.7,0) circle [radius=0.05];
\draw[fill] (6.2,0) circle [radius=0.05];
\draw[fill] (5.4,0) circle [radius=0.05];
\draw[fill] (3.6,0) circle [radius=0.05];
\draw[fill] (2.8,0) circle [radius=0.05];
\draw[arrow][thick](4.5,2) arc (90:0:2);
\end{tikzpicture}}
\end{center}
\end{figure}
\centerline{\small Figure 1\quad The branch cuts and points.}
\vspace{0.2cm}

%\centerline{\includegraphics[scale=1.2]{actmark.eps}}
%\centerline{\small Figure 1\quad Journal mark}

%\noindent {\small\textbf{Figure 1.} The branch cuts and points are expressed by two blue segments and four black points, respectively. The pink area denotes $\Sigma$.}
%

According to the above diagram, we have
\begin{equation}\nonumber
\begin{aligned}
&\Sigma_{+}=[1,(r +A)]\cup[-(r + A),-1],\notag\\
&\Sigma_{-}=[(r-A),1]\cup[-1,-(r-A)].
\end{aligned}
\end{equation}
%that is
%\begin{align}\nonumber
%\Sigma=\Sigma_{+}\cup\Sigma_{-}.
%\end{align}
Let
\begin{align}\nonumber
\Omega_{0}:=\{z:|z|=1\},~~\Omega_{in}:=\{z:|z|<1\},~~\Omega_{out}:=\{z:1<|z|<\infty\}.
\end{align}
We obtain
\begin{equation}\label{AF-7-1}
\begin{split}
&|\zeta(z)|=1,~~z\in\Omega_{0}\cup\Sigma,\\
&|\xi(z)|=1,~~z\in\Sigma.
\end{split}
\end{equation}
In the piece $M_{+}$, the meromorphic functions $\zeta(z)$ and $\xi(z)$ have the removable singularity at $z=\infty$ and the first order pole at $z=0$.
In the piece $M_{-}$, the analytic functions $\zeta(z)$ and $\xi(z)$ possess the removable singularity at $z=0$ and the first order pole at $z=\infty$.
For the sake of simplicity, we only need to analyze one piece, taking $M_{-}$ as an example.
%the piece $M_{-}$ since we can apply the similar analysis for the piece $M_{+}$.
Considering the analytic of $\zeta(z)$ and $\xi(z)$ in the region $\Omega_{in}\setminus\Sigma_{-}$, and they satisfy \eqref{AF-7-1} when $z\in\Sigma_{-}$, we know that
\begin{align}\nonumber
|\xi(z)|\leq \max\{1,\max_{|\lambda|=1}|\xi(z)|\}=(1+r)/A.
\end{align}
In view of the maximum modulus principle, one has $|\zeta(z)|\leq 1$.

We then give a new gauge transformation
\begin{equation}\nonumber
\begin{aligned}
&w_{n}^{+}=\left(\prod_{k=n}^{+\infty}\frac{1+|v_{k}|^{2}}{r^{2}}\right)
T^{-1}\chi_{n}^{+},~~w_{n}^{-}=T^{-1}\chi_{n}^{-}.
\end{aligned}
\end{equation}
Combining the NZBCs \eqref{AF-5} and Eq. \eqref{AF-7}, we obtain
\begin{equation}\label{AF-10}
\begin{split}
&\varphi_{n}^{+}=T-r^{-2}\sum_{k=n}^{+\infty}
T(r\zeta^{\sigma_{3}})^{n-k}T^{-1}N_{k}^{+}\varphi_{k+1}^{+}(r\zeta^{\sigma_{3}})^{-n+k+1},\\
&\varphi_{n}^{-}=T+\sum_{k=-\infty}^{n-1}
T(r\zeta^{\sigma_{3}})^{n-(k+1)}T^{-1}N_{k}^{-}\varphi_{k}^{-}(r\zeta^{\sigma_{3}})^{-n+k},
\end{split}
\end{equation}
in which
\begin{align}\label{AF-8}
\varphi_{n}^{\pm}=Tw_{n}^{\pm}(r\zeta^{\sigma_{3}})^{-n}=\chi_{n}^{\pm}(r\zeta^{\sigma_{3}})^{-n}.
\end{align}
For solutions $\varphi_{n}^{\pm}=[\varphi_{n,1}^{\pm},\varphi_{n,2}^{\pm}]$, it follows from Eq. \eqref{AF-10} that we have
\begin{equation}\label{AF-11}
\begin{split}
&\varphi_{n,1}^{+}=\left(
               \begin{array}{c}
                 1 \\
                 \xi \\
               \end{array}
             \right)-\zeta\sum_{k=n}^{+\infty}
E_{1}(k,n,z)r^{-1}N_{k}^{+}\varphi_{k+1,1}^{+},\\
&\varphi_{n,1}^{-}=\left(
               \begin{array}{c}
                 1 \\
                 \xi \\
               \end{array}
             \right)+\zeta^{-1}\sum_{k=-\infty}^{n-1}
E_{1}(k+1,n,z)r^{-1}N_{k}^{-}\varphi_{k,1}^{-},\\
&\varphi_{n,2}^{+}=\left(
               \begin{array}{c}
                 \xi \\
                 1 \\
               \end{array}
             \right)-\zeta^{-1}\sum_{k=n}^{+\infty}
E_{2}(k,n,z)r^{-1}N_{k}^{+}\varphi_{k+1,2}^{+},\\
&\varphi_{n,2}^{-}=\left(
               \begin{array}{c}
                 \xi \\
                 1 \\
               \end{array}
             \right)+\zeta\sum_{k=-\infty}^{n-1}
E_{2}(k+1,n,z)r^{-1}N_{k}^{-}\varphi_{k,2}^{-},
\end{split}
\end{equation}
where
\begin{equation}\nonumber
\begin{aligned}
&E_{1}(k,n,z)=T\left(
     \begin{array}{cc}
       1 & 0 \\
       0 & \zeta^{2(k-n)} \\
     \end{array}
   \right)T^{-1}=\mathbb{I}+\frac{\zeta^{2(k-n)}-1}{1-\xi^{2}}\left(
                                                       \begin{array}{cc}
                                                         -\xi^{2} & \xi \\
                                                         -\xi & 1 \\
                                                       \end{array}
                                                     \right)
   ,\notag\\
&E_{2}(k,n,z)=T\left(
     \begin{array}{cc}
       \zeta^{-2(k-n)} & 0 \\
       0 & 0 \\
     \end{array}
   \right)T^{-1}=\mathbb{I}+\frac{\zeta^{-2(k-n)}-1}{1-\xi^{2}}\left(
                                                       \begin{array}{cc}
                                                         1 & -\xi \\
                                                         \xi & -\xi^{2} \\
                                                       \end{array}
                                                     \right).
\end{aligned}
\end{equation}
\vspace{0.2cm}

\noindent
\textbf{Proposition 2.1}
For any finite integer $n_{0}$, when $\sum_{k=n_{0}}^{\pm\infty}(1+|k|)|v_{k}\textrm{e}^{-\textrm{i}B_{\pm}}-A|<\pm\infty$, the solutions $\varphi_{n,1}^{+}$ and $\varphi_{n,2}^{-}$ are analytic in the region $\Omega_{in}\setminus\Sigma_{-}$ and continuous to its boundary; $\varphi_{n,1}^{-}$ and
$\varphi_{n,2}^{+}$ are analytic in the region $\Omega_{out} \setminus\Sigma_{+}$ and continuous to its boundary.
\vspace{0.2cm}

\begin{proof}
The proof is similar to Reference \cite{P-1979}.
\end{proof}

The Jost solutions
\begin{align}\label{AF-10-1}
J_{\pm}(n,z)=\psi_{n}^{\pm}(M(z))^{-n}
\end{align}
are linear dependent, and their relationship is expressed as
\begin{align}\label{AF-12}
J_{-}(n,z)=J_{+}(n,z)M(z)^{n}S(z)(M(z))^{-n},~~z\in\Omega_{0},
\end{align}
through the scattering matrix
\begin{align}\label{AF-13}
S(z)=\left(
             \begin{array}{cc}
               a(z) & c(z) \\
               b(z) & d(z) \\
             \end{array}
           \right),
\end{align}
and $M(z)=\mathrm{diag}\left(r\zeta(z), r\zeta(z)^{-1}\right)$.

It follows from Eqs. \eqref{AF-4} and \eqref{AF-10-1} that one has
\begin{align}\label{AF-14}
J_{\pm}(n+1,z)=X_{n}(z)J_{\pm}(n,z)M(z)^{-1}.
\end{align}
Furthermore, we have
\begin{align}\nonumber
J_{\pm}^{\dagger}(n+1,z^{\ast})=[M(z^{\ast})^{-1}]^{\dagger}J_{\pm}^{\dagger}(n,z^{\ast})
X_{n}^{\dagger}(z^{\ast})=r^{-2}M(z)J_{\pm}^{\dagger}(n,z^{\ast})
X_{n}^{\dagger}(z^{\ast}),
\end{align}
where $z^{\ast}\equiv\bar{z}^{-1}$, and $\dagger$ denotes the Hermite conjugate.

Let $A_{n}=1+|v_{n}|^2$, then $X_{n}^{\dagger}(z^{\ast})X_{n}(z)=A_{n}\mathbb{I}$,
we compute
\begin{equation}\nonumber
\begin{aligned}
J_{\pm}^{\dagger}(n+1,z^{\ast})J_{\pm}(n+1,z)
=A_{n}r^{-2}M(z)J_{\pm}^{\dagger}(n,z^{\ast})
J_{\pm}(n,z)M(z)^{-1}.
\end{aligned}
\end{equation}
The forms of boundary conditions of $J_{\pm}(n,z)$ and $J_{\pm}^{\dagger}(n,z^{\ast})$ are as follows
\begin{equation}\nonumber
\begin{aligned}
&J_{\pm}(\pm\infty,z)
=\textrm{e}^{\frac{\textrm{i}}{2}B_{\pm}\sigma_{3}}\left(
                                           \begin{array}{cc}
                                             1 & A^{-1}(r\zeta-z) \\
                                             A^{-1}(r\zeta-z) & 1 \\
                                           \end{array}
                                         \right),\notag\\
&J_{\pm}^{\dagger}(\pm\infty,z^{\ast})=\left(
                                           \begin{array}{cc}
                                             1 & A^{-1}(r\zeta^{-1}-z^{-1}) \\
                                             A^{-1}(r\zeta^{-1}-z^{-1}) & 1 \\
                                           \end{array}
                                         \right)\textrm{e}^{-\frac{\textrm{i}}{2}B_{\pm}\sigma_{3}}.
\end{aligned}
\end{equation}
Similar to Proposition 2.1, we have the following analyticity of the Jost solutions $J_{\pm}$.
\vspace{0.2cm}

\noindent
\textbf{Proposition 2.2}
For any finite integer $n_{0}$, when $\sum_{k=n_{0}}^{\pm\infty}(1+|k|)|v_{k}\textrm{e}^{-\textrm{i}B_{\pm}}-A|<\pm\infty$, the solutions $J_{+,1}$ and $J_{-,2}$ are analytic in the region $\Omega_{in}\setminus\Sigma_{-}$; $J_{-,1}$ and
$J_{+,2}$ are analytic in the region $\Omega_{out} \setminus\Sigma_{+}$.
\vspace{0.2cm}

\begin{proof}
With the aid of Proposition 2.1 and the definitions of $\psi_{n}^{\pm}$, $\varphi_{n}^{\pm}$ and $J_{\pm}$ that are given by \eqref{AF-6}, \eqref{AF-8} and \eqref{AF-10-1}, one can establish this proposition.
\end{proof}
\vspace{0.2cm}

\noindent
\textbf{Proposition 2.3}
The Jost solutions $J_{\pm}$ satisfy
\begin{equation}\nonumber
\begin{aligned}
&J_{+}^{\dagger}(n,z^{\ast})J_{+}(n,z)
=(1-\xi^{2})\prod_{l=n}^{+\infty}\left(A_{l}^{-1}r^{2}\right)\mathbb{I}=(1-\xi^{2})\Gamma_{n}^{+},\notag\\
&J_{-}^{\dagger}(n,z^{\ast})J_{-}(n,z)
=(1-\xi^{2})\prod_{l=-\infty}^{n-1}\left(A_{l}r^{-2}\right)\mathbb{I}=(1-\xi^{2})\Gamma_{n}^{-}.
\end{aligned}
\end{equation}
\vspace{0.2cm}

Based on the Proposition 2.3, we arrive at
\begin{equation}\nonumber
\begin{aligned}
&\det(J_{+}(n,z))
=(1-\xi^{2})\prod_{l=n}^{+\infty}r^{2}A_{l}^{-1},\notag\\
&\det(J_{-}(n,z))
=(1-\xi^{2})\prod_{l=-\infty}^{n-1}r^{-2}A_{l}.
\end{aligned}
\end{equation}
From \eqref{AF-12}, we deduce
\begin{equation}\nonumber
\begin{aligned}
&\det(S(z))
=\prod_{l=-\infty}^{+\infty}r^{2}A_{l}^{-1}\equiv \phi,\notag\\
&J_{-}^{\dagger}(n,z^{\ast})=M(z)^{n}S^{\dagger}(z^{\ast})(M(z))^{-n}
J_{+}^{\dagger}(n,z^{\ast}),\notag\\
&J_{-}^{-1}(n,z)=M(z)^{n}S^{-1}(z)(M(z))^{-n}J_{+}^{-1}(n,z).
\end{aligned}
\end{equation}
Considering the symmetry
\begin{equation}\nonumber
\begin{aligned}
S^{\dagger}(z^{\ast})=\phi S(z)^{-1},
\end{aligned}
\end{equation}
the scattering matrix $S(z)$ is written as
\begin{align}\label{AF-15}
S(z)=\left(
             \begin{array}{cc}
               a(z) & -\bar{b}(z^{\ast}) \\
               b(z) & \bar{a}(z^{\ast}) \\
             \end{array}
           \right),
\end{align}
and the scattering coefficient is $r(z)=b(z)a(z)^{-1}$.

In view of the Jost solutions $J_{\pm}(n,z)$ satisfying
\begin{align}\label{AF-16}
J_{\pm}(n,z)=\sigma_{3}J_{\pm}(n,-z)\sigma_{3},
\end{align}
so we conclude
\begin{align}\label{AF-16-1}
S(z)=\sigma_{3}S(-z)\sigma_{3}.
\end{align}
Furthermore, it follows from Proposition 2.2 that we have the sectionally analytic matrix function
\begin{equation}\nonumber
\begin{aligned}
\Phi(n,z)=\left\{\begin{array}{c}
              (J_{-,1}(n,z),J_{+,2}(n,z)),~~\lambda\in\Omega_{out}\setminus\Sigma_{+},\\
              (J_{+,1}(n,z),J_{-,2}(n,z)),~~\lambda\in\Omega_{in}\setminus\Sigma_{-}.
            \end{array}
\right.
\end{aligned}
\end{equation}
\vspace{0.2cm}

\noindent
\textbf{Proposition 2.4}
The asymptotic behaviors of matrices $\Phi^{\pm}(n,z)$ are as follows
\begin{equation}\label{AF-17}
\begin{split}
&\Phi^{+}(n,z)=\left(
                       \begin{array}{cc}
                         \textrm{e}^{\frac{\textrm{i}}{2}B_{-}} & 0 \\
                         0 & \textrm{e}^{-\frac{\textrm{i}}{2}B_{+}}\Gamma_{n}^{+} \\
                       \end{array}
                     \right)+O(z^{-1}),~~z\rightarrow\infty,\\
&\Phi^{-}(n,z)=\left(
                       \begin{array}{cc}
                         \textrm{e}^{\frac{\textrm{i}}{2}B_{+}}\Gamma_{n}^{-} & 0 \\
                         0 & \textrm{e}^{-\frac{\textrm{i}}{2}B_{-}} \\
                       \end{array}
                     \right)+O(z),~~~~~z\rightarrow0.
\end{split}
\end{equation}
\vspace{0.2cm}

It follows from \eqref{AF-12} and \eqref{AF-15} that we have
\begin{equation}\label{AF-18}
\begin{split}
a(z)&=\frac{\det((J_{-,1}(n,z),J_{+,2}(n,z)))}{(1-\xi^{2})\Gamma_{n}^{+}},\\
\bar{a}(z^{\ast})&=\frac{\det((J_{+,1}(n,z),J_{-,2}(n,z)))}{(1-\xi^{2})\Gamma_{n}^{+}},\\
b(z)&=\frac{\det((J_{+,1}(n,z),J_{-,1}(n,z)))}{\zeta^{-2n}(1-\xi^{2})\Gamma_{n}^{+}},\\
\bar{b}(z^{\ast})&=\frac{\det((J_{-,2}(n,z),J_{+,2}(n,z)))}{\zeta^{2n}(1-\xi^{2})\Gamma_{n}^{+}}.
\end{split}
\end{equation}
In terms of \eqref{AF-17}, the asymptotic expression of $a(z)$ is
\begin{align}\nonumber
a(z)=\textrm{e}^{\frac{\textrm{i}}{2}(B_{-}-B_{+})}+O\left(\frac{1}{z}\right),~~z\rightarrow\infty.
\end{align}
The function $a(z)$ has analytic continuation when $z\in\Omega_{out}\setminus\Sigma_{+}$. For the sake of simplicity, suppose that it only possesses finite zeros, all of which are inside the boundary $\partial(\Omega_{out}\setminus\Sigma_{+})$, then we have
\begin{align}\label{AF-18-1}
a(z)=\hat{a}(z)\prod_{i=1}^{k}\left(\frac{z^{2}-z_{i}^{2}}{z^{2}
-(z_{i}^{\ast})^{2}}\right)^{m_{i}},
\end{align}
here the analytic function $\hat{a}(z)$ does not have zero points in the region $ \Omega_{out}\setminus\Sigma_{+}$, and the orders of zeros are denoted by $m_{i}$ which belongs to $\mathbb{Z}^{+}$.

Define
\begin{align}\nonumber
\Phi_{\pm}(n,z)=\Phi^{\pm}(n,z)\zeta^{n\sigma_{3}},
\end{align}
and
\begin{align}\nonumber
\Phi_{+}(n,z)=\sum_{j=0}^{\infty}\Phi_{+}^{[j]}(n,z_{i})(z-z_{i})^{j}.
\end{align}
$\mathrm{Ker}(\Phi_{+}(n,z))$ at $z=z_{i}$ is
\begin{align}\label{AF-19}
\left(
\begin{array}{cccc}
\Phi_{+}^{[0]}(n,z_{i}) & 0 & \cdots & 0 \\
\Phi_{+}^{[1]}(n,z_{i}) & \Phi_{+}^{[0]}(n,z_{i}) & \cdots & 0 \\
\vdots & \vdots & \ddots & \vdots \\
\Phi_{+}^{[m_{i}-1]}(n,z_{i}) & \Phi_{+}^{[m_{i}-2]}(n,z_{i}) & \cdots & \Phi_{+}^{[0]}(n,z_{i}) \\
\end{array}
\right)\left(
         \begin{array}{c}
           \omega_{0}(z_{i}) \\
           \omega_{1}(z_{i}) \\
           \vdots \\
           \omega_{m_{i}-1}(z_{i}) \\
         \end{array}
       \right)=0.
\end{align}
Moreover, we have
\begin{align}\label{AF-20}
\bar{\Phi}_{+}(n,z)=\sigma_{2}\Phi_{-}(n,z^{\ast})\sigma_{2},~~\sigma_{2}=\left(
                                                                                        \begin{array}{cc}
                                                                                          0 & -\textrm{i} \\
                                                                                          \textrm{i} & 0 \\
                                                                                        \end{array}
                                                                                      \right),
\end{align}
and
\begin{align}\label{AF-21}
\left(
\begin{array}{cccc}
\Phi_{-}^{[0]}(n,z_{i}^{\ast}) & 0 & \cdots & 0 \\
\Phi_{-}^{[1]}(n,z_{i}^{\ast}) & \Phi_{-}^{[0]}(n,z_{i}^{\ast}) & \cdots & 0 \\
\vdots & \vdots & \ddots & \vdots \\
\Phi_{-}^{[m_{i}-1]}(n,z_{i}^{\ast}) & \Phi_{-}^{[m_{i}-2]}(n,z_{i}^{\ast}) & \cdots & \Phi_{-}^{[0]}(n,z_{i}^{\ast}) \\
\end{array}
\right)\left(
         \begin{array}{c}
           \sigma_{2}\overline{\omega_{0}(z_{i})} \\
           \sigma_{2}\overline{\omega_{1}(z_{i})} \\
           \vdots \\
           \sigma_{2}\overline{\omega_{m_{i}-1}(z_{i})} \\
         \end{array}
       \right)=0,
\end{align}
here
\begin{align}\nonumber
\Phi_{-}(n,z)=\sum_{j=0}^{\infty}\Phi_{-}^{[j]}(n,z_{i}^{\ast})(z-z_{i}^{\ast})^{j}.
\end{align}
According to \eqref{AF-16}, $\mathrm{Ker}(\Phi_{+}(n,z))$ at $z=-z_{i}$ is
\begin{align}\label{AF-22}
\left(
\begin{array}{cccc}
\Phi_{+}^{[0]}(n,-z_{i}) & 0 & \cdots & 0 \\
\Phi_{+}^{[1]}(n,-z_{i}) & \Phi_{+}^{[0]}(n,-z_{i}) & \cdots & 0 \\
\vdots & \vdots & \ddots & \vdots \\
\Phi_{+}^{[m_{i}-1]}(n,-z_{i}) & \Phi_{+}^{[m_{i}-2]}(n,-z_{i}) & \cdots & \Phi_{+}^{[0]}(n,-z_{i}) \\
\end{array}
\right)\left(
         \begin{array}{c}
           \sigma_{3}\omega_{0}(z_{i}) \\
           \sigma_{3}\omega_{1}(z_{i}) \\
           \vdots \\
           \sigma_{3}\omega_{m_{i}-1}(z_{i}) \\
         \end{array}
       \right)=0,
\end{align}
with
\begin{align}\nonumber
\Phi_{+}(n,z)=\sum_{j=0}^{\infty}\Phi_{+}^{[j]}(n,-z_{i})(z+z_{i})^{j}.
\end{align}
$\mathrm{Ker}(\Phi_{-}(n,z))$ at $z=-z_{i}^{\ast}$ is
\begin{align}\label{AF-23}
\left(
\begin{array}{cccc}
\Phi_{-}^{[0]}(n,-z_{i}^{\ast}) & 0 & \cdots & 0 \\
\Phi_{-}^{[1]}(n,-z_{i}^{\ast}) & \Phi_{-}^{[0]}(n,-z_{i}^{\ast}) & \cdots & 0 \\
\vdots & \vdots & \ddots & \vdots \\
\Phi_{-}^{[m_{i}-1]}(n,-z_{i}^{\ast}) & \Phi_{-}^{[m_{i}-2]}(n,-z_{i}^{\ast}) & \cdots & \Phi_{-}^{[0]}(n,-z_{i}^{\ast}) \\
\end{array}
\right)\left(
         \begin{array}{c}
           \sigma_{3}\sigma_{2}\overline{\omega_{0}(z_{i})} \\
           \sigma_{3}\sigma_{2}\overline{\omega_{1}(z_{i})} \\
           \vdots \\
           \sigma_{3}\sigma_{2}\overline{\omega_{m_{i}-1}(z_{i})} \\
         \end{array}
       \right)=0,
\end{align}
with
\begin{align}\nonumber
\Phi_{-}(n,z)=\sum_{j=0}^{\infty}\Phi_{-}^{[j]}(n,-z_{i}^{\ast})(z+z_{i}^{\ast})^{j}.
\end{align}
For meromorphic functions $\Phi_{\pm}(n,z)$, the degenerate properties in the neighborhood of $\pm z_{i}$ and $\pm z_{i}^{\ast}$ are given by \eqref{AF-19}, \eqref{AF-21}, \eqref{AF-22}, \eqref{AF-23}, respectively. Further on, the rational solutions can be constructed.

\subsection{Riemann-Hilbert problem 1}
In order to construct the corresponding RHP, we introduce the sectional meromorphic function
\begin{align}\label{AF-24}
N(n,z)=\left\{\begin{array}{cc}
               \left(
                                  \begin{array}{cc}
                                    J_{-,1}(n,z)a(z)^{-1} & J_{+,2}(n,z) \\
                                  \end{array}
                                \right),
                & z\in\Omega_{out}\setminus\Sigma_{+}, \\
               \left(
                                  \begin{array}{cc}
                                    J_{+,1}(n,z) & J_{-,2}(n,z)\bar{a}(z^{\ast})^{-1} \\
                                  \end{array}
                                \right),
               & z\in\Omega_{in}\setminus\Sigma_{-}.
             \end{array}
\right.
\end{align}
$N^{\pm}(n,z)$ can be rewritten as
\begin{equation}\nonumber
\begin{aligned}
&N^{+}(n,z)=J_{+}(n,z)\zeta^{n\sigma_{3}}\left(
                                                       \begin{array}{cc}
                                                         1 & 0 \\
                                                         r(z) & 1 \\
                                                       \end{array}
                                                     \right)\zeta^{-n\sigma_{3}},\notag\\
&N^{-}(n,z)=J_{+}(n,z)\zeta^{n\sigma_{3}}\left(
                                                       \begin{array}{cc}
                                                         1 & -\bar{r}(z^{\ast}) \\
                                                         0 & 1 \\
                                                       \end{array}
                                                     \right)\zeta^{-n\sigma_{3}}.
\end{aligned}
\end{equation}
Therefore the relationship between $N^{+}(n,z)$ and $N^{-}(n,z)$ is expressed as
\begin{align}\nonumber
N^{+}(n,z)=N^{-}(n,z)G_{1},~~z\in\Omega,
\end{align}
by the jump matrix
\begin{align}\label{AF-25}
G_{1}=\zeta^{n\sigma_{3}}\left(
                                                       \begin{array}{cc}
                                                         1+r(z)\bar{r}(z^{\ast}) & \bar{r}(z^{\ast}) \\
                                                         r(z) & 1 \\
                                                       \end{array}
                                                     \right)\zeta^{-n\sigma_{3}},~~z\in\Omega.
\end{align}
When $z\in\Sigma$, the jump condition needs to be explored. The boundary conditions for $J_{\pm}(n,z)$ are
\begin{align}\nonumber
J_{\pm}(\pm\infty,z_{\pm})=\textrm{e}^{\frac{\textrm{i}}{2}B_{\pm}\sigma_{3}}\left(
                                                                        \begin{array}{cc}
                                                                          1 & \xi_{\pm} \\
                                                                          \xi_{\pm} & 1 \\
                                                                        \end{array}
                                                                      \right),
\end{align}
where
\begin{align}\nonumber
z_{\pm}=z\pm \textrm{i}\epsilon,~~\epsilon\rightarrow0^{+},~~\xi_{\pm}=\xi(z_{\pm}).
\end{align}
Thus, we calculate
\begin{align}\nonumber
J_{\pm}(\pm\infty,z_{+})=J_{\pm}(\pm\infty,z_{-})S^{'},~~z\in\Sigma,
\end{align}
with
\begin{align}\nonumber
S^{'}=\left(
        \begin{array}{cc}
          0 & \xi_{+} \\
          \xi_{+} & 0 \\
        \end{array}
      \right).
\end{align}
On the basis of uniqueness of solutions of difference equations, one concludes
\begin{align}\nonumber
J_{\pm}(n,z_{+})=J_{\pm}(n,z_{-})S^{'},~~z\in\Sigma.
\end{align}
From \eqref{AF-18}, we compute
\begin{equation}\nonumber
\begin{aligned}
&a(z_{+})=\frac{\det((J_{-,1}(n,z_{+}),J_{+,2}(n,z_{+})))}{(1-\xi_{+}^{2})\Gamma_{n}}\notag\\
&~~~~~~~~=\frac{\xi_{+}^{2}(1-\xi_{-}^{2})}{(1-\xi_{+}^{2})}\frac{\det((J_{-,2}(n,z_{-}),J_{+,1}(n,z_{-})))}
{(1-\xi_{-}^{2})\Gamma_{n}}\notag\\
&~~~~~~~~=\bar{a}(z_{-}^{\ast}),\notag\\
&b(z_{+})=\frac{\det((J_{+,1}(n,z_{+}),J_{-,1}(n,z_{+})))}{\xi_{+}^{-2n}(1-\xi_{+}^{2})\Gamma_{n}}\notag\\
&~~~~~~~~=\frac{\xi_{+}^{2}(1-\xi_{-}^{2})}{(1-\xi_{+}^{2})}\frac{\det((J_{+,2}(n,z_{-}),J_{-,2}(n,z_{-})))}
{\xi_{+}^{-2n}(1-\xi_{-}^{2})\Gamma_{n}}\notag\\
&~~~~~~~~=-\bar{b}(z_{-}^{\ast}),\notag\\
&r(z_{+})=-\bar{r}(z_{-}^{\ast}).
\end{aligned}
\end{equation}
When $z\in\Sigma_{+}$ and $z\in\Sigma_{-}$, we severally obtain the jump conditions
\begin{equation}\nonumber
\begin{aligned}
N^{+}(n,z_{+})&=J_{+}(n,z_{+})\zeta_{+}^{n\sigma_{3}}\left(
                                                       \begin{array}{cc}
                                                         1 & 0 \\
                                                         r(z_{+}) & 1 \\
                                                       \end{array}
                                                     \right)\zeta_{+}^{-n\sigma_{3}}\notag\\
&=N^{+}(n,z_{-})\xi_{-}^{-1}\left(
                                                       \begin{array}{cc}
                                                         1 & 0 \\
                                                         r(z_{-})\zeta_{-}^{-2n} & 1 \\
                                                       \end{array}
                                                     \right)\left(
                                                              \begin{array}{cc}
                                                                0 & 1 \\
                                                                1 & 0 \\
                                                              \end{array}
                                                            \right)\left(
                                                       \begin{array}{cc}
                                                         1 & 0 \\
                                                         r(z_{+})\zeta_{+}^{-2n} & 1 \\
                                                       \end{array}
                                                     \right)\notag\\
&=N^{+}(n,z_{-})G_{2,+},\notag\\
N^{+}(n,z_{+})&=N^{+}(n,z_{-})G_{2,-},
\end{aligned}
\end{equation}
where
\begin{equation}\label{AF-26}
\begin{split}
&G_{2,+}=\xi_{-}^{-1}\left(
                       \begin{array}{cc}
                         -\bar{r}(z_{-}^{\ast})\zeta_{-}^{2n} & 1 \\
                         1+r(z_{-})\bar{r}(z_{-}^{\ast}) & -r(z_{-})\zeta_{-}^{-2n} \\
                       \end{array}
                     \right),\\
&G_{2,-}=\xi_{-}^{-1}\left(
                       \begin{array}{cc}
                         \bar{r}(z_{-}^{\ast})\zeta_{-}^{2n} & 1+r(z_{-})\bar{r}(z_{-}^{\ast}) \\
                         1 & r(z_{-})\zeta_{-}^{-2n} \\
                       \end{array}
                     \right).
\end{split}
\end{equation}
In conclusion, we have the following definition of RHP 1.
\vspace{0.2cm}

\noindent
\textbf{Definition 2.5}
The function $N(n,z)$ satisfies that
\begin{itemize}
  \item it is sectionally analytic as $z\in\mathbb{C}\setminus\left\{\Omega\cup\Sigma\right\}$;
  \item the jump relations
  \begin{equation}\nonumber
\begin{aligned}
N^{+}(n,z)=\left\{\begin{array}{cc}
                          N^{-}(n,z)G_{1},~ & z\in \Omega,~ \\
                          N^{-}(n,z)G_{2,+}, & z\in \Sigma_{+}, \\
                          N^{-}(n,z)G_{2,-}, & z\in \Sigma_{-},
                        \end{array}
\right.
\end{aligned}
\end{equation}
where $G_{1}$ and $G_{2,\pm}$ are given by \eqref{AF-25} and \eqref{AF-26};
  \item the dominating part of $N(n,z)$ is
\begin{equation}\nonumber
\begin{aligned}
N(n,z)=\left[\sum_{i=1}^{k}\sum_{j=1}^{m_{i}}\left(\frac{N_{1,j}^{[i]}(n)}{(z-z_{i})^{j}}
+\frac{N_{1,j}^{[-i]}(n)}{(z+z_{i})^{j}}\right),\sum_{i=1}^{n}\sum_{j=1}^{m_{i}}
\left(\frac{N_{2,j}^{[i]}(n)}{(z-z_{i}^{\ast})^{j}}
+\frac{N_{2,j}^{[-i]}(n)}{(z+z_{i}^{\ast})^{j}}\right)\right],
\end{aligned}
\end{equation}
where $N_{k,j}^{[s]}$ denote the column vectors, and are defined by the analytic part of $N(n,z)$ with conditions \eqref{AF-19}, \eqref{AF-21}, \eqref{AF-22}, \eqref{AF-23};
  \item the normalization conditions
\begin{equation}\nonumber
\begin{aligned}
&N(n,z)=\left(
                       \begin{array}{cc}
                         \textrm{e}^{\frac{\textrm{i}}{2}B_{+}}\Gamma_{n}^{-} & 0 \\
                         0 & \textrm{e}^{-\frac{\textrm{i}}{2}B_{-}} \\
                       \end{array}
                     \right)+O(z),~~~~~z\rightarrow0,\notag\\
&N(n,z)=\left(
                       \begin{array}{cc}
                         \textrm{e}^{\frac{\textrm{i}}{2}B_{-}} & 0 \\
                         0 & \textrm{e}^{-\frac{\textrm{i}}{2}B_{+}}\Gamma_{n}^{+} \\
                       \end{array}
                     \right)+O\left(\frac{1}{z}\right),~~z\rightarrow\infty.
\end{aligned}
\end{equation}
\end{itemize}
\vspace{0.2cm}

According to the argument of Zhou's vanishing lemma in \cite{X-1989}, we obtain the existence and uniqueness of solutions of the sectional function $N(n,z)$ in \eqref{AF-24} by changing the poles into jumps along small circular contours as well as the Schwartz reflection about $\Omega_{0}$.

\subsection{Scattering data}

\noindent
\textbf{Lemma 2.6}
The requirement on $v_{n}(t)$ is that it satisfies Eq. \eqref{AF-3} and $v_{n}(t)\rightarrow A \textrm{e}^{\textrm{i}B_{\pm}}$ when $n\rightarrow\pm\infty$. For $\forall$ $t\in [0, +\infty)$, $\psi^{\pm}_{n}(t,z)$ are the Jost solutions. Then
\begin{align}\nonumber
\eta_{\pm}(n, t,z) = \psi^{\pm}_{n}(t,z)\mu^{\pm}(t,z)
\end{align}
deal with the Lax pair \eqref{AF-4} synchronously, in which
\begin{equation}\label{AF-27}
\begin{split}
\mu^{\pm}(t,z)=&\textrm{e}^{\left[A^{2}(a\sin B+b\cos B)+\delta(z)\omega\sigma_{3}\right]t},\\
\delta(z):=&(b+\textrm{i}a)z^{-1}\textrm{e}^{-\textrm{i}B}+(b-\textrm{i}a)z\textrm{e}^{\textrm{i}B},\\
\omega(z)=&\frac{\sqrt{(1+z^{2})^{2}-4r^{2}z^{2}}}{2z}.
\end{split}
\end{equation}
\vspace{0.2cm}

\begin{proof}
Given the fixed $t$, the functions $\psi^{\pm}_{n}(t,z)$ solve the $n$-part of Lax pair, so there exist solutions
\begin{align}\nonumber
\eta_{\pm}(n, t,z) = \psi^{\pm}_{n}(t,z)\mu^{\pm}(t,z)
\end{align}
deal with the Lax pair \eqref{AF-4} simultaneously on the basis of the fundamental solution theory of difference equation. Then we obtain
\begin{align}\label{AF-27-0}
\frac{d\psi^{\pm}_{n}(t,z)}{d t}\mu^{\pm}(t,z)+\psi^{\pm}_{n}(t,z)\frac{d\mu^{\pm}(t,z)}
{d t}=T_{n}\psi^{\pm}_{n}(t,z)\mu^{\pm}(t,z).
\end{align}
It follows from \eqref{AF-5} that
\begin{equation}\nonumber
\begin{aligned}
&\psi^{\pm}_{n}(t,z)\rightarrow\textrm{e}^{\frac{\textrm{i}}{2}B_{\pm}\sigma_{3}}\left(
                                                                          \begin{array}{cc}
                                                                            1 & \xi \\
                                                                            \xi & 1 \\
                                                                          \end{array}
                                                                        \right)r^{n}\zeta^{n\sigma_{3}}
,\notag\\
&T_{n}\rightarrow T^{\pm}_{n}=\textrm{e}^{\frac{\textrm{i}}{2}B_{\pm}\hat{\sigma}_{3}}\left(
              \begin{array}{cc}
                z\delta(z)+D(z) & A\delta(z) \\
                -A\delta(z) & z^{-1}\delta(z)+D(z) \\
              \end{array}
            \right),
\end{aligned}
\end{equation}
where
\begin{equation}\label{AF-27-1}
\begin{split}
\textrm{e}^{a\hat{\sigma}_{3}}\mathcal{A}=&\textrm{e}^{a\sigma_{3}}\mathcal{A}\textrm{e}^{-a\sigma_{3}},\\
D(z)=&\frac{\textrm{i}a}{2}(\textrm{e}^{\textrm{i}B}z^{2}-\textrm{e}^{-\textrm{i}B}z^{-2})+\textrm{i}
(ar^{2}\cos B-b(1+A^{2})\sin B)\\
&+(b-\textrm{i}a)A^{2}\textrm{e}^{\textrm{i}B}-(b+\textrm{i}a)\textrm{e}^{-\textrm{i}B}.
\end{split}
\end{equation}
Based on \eqref{AF-27-0} and the initial conditions
\begin{align}\nonumber
\mu^{\pm}(0,z)= \mathbb{I}_{2},
\end{align}
the solutions $\mu^{\pm}(t,z)$ in \eqref{AF-27} are calculated. This proves the lemma.
\end{proof}

We obtain the evolution form of scattering data
\begin{equation}\nonumber
\begin{aligned}
&\eta_{+}(n,t,z)=\eta_{-}(n,t,z)S(z),\notag\\
&\eta_{\pm}(n,t,z)=\Phi_{\pm}(n,t,z)\mu^{\pm}(t,z)
\end{aligned}
\end{equation}
in light of the evolution of Jost solutions, and the expression of $\mathrm{Ker}(\Phi_{\pm}(n,z))$ at points $z_{i}$ $(i=1,2,\ldots,k)$ is
\begin{align}\nonumber
\left(
\begin{array}{cccc}
\eta_{+}^{[0]}(n,t,z_{i}) & 0 & \cdots & 0 \\
\eta_{+}^{[1]}(n,t,z_{i}) & \eta_{+}^{[0]}(n,t,z_{i}) & \cdots & 0 \\
\vdots & \vdots & \ddots & \vdots \\
\eta_{+}^{[m_{i}-1]}(n,t,z_{i}) & \eta_{+}^{[m_{i}-2]}(n,t,z_{i}) & \cdots & \eta_{+}^{[0]}(n,t,z_{i}) \\
\end{array}
\right)\left(
         \begin{array}{c}
           \omega_{0}(z_{i}) \\
           \omega_{1}(z_{i}) \\
           \vdots \\
           \omega_{m_{i}-1}(z_{i}) \\
         \end{array}
       \right)=0,
\end{align}
with
\begin{align}\nonumber
\eta_{+}(n,t,z)=\sum_{j=0}^{\infty} \eta_{+}^{[j]}(n,t,z_{i})(z-z_{i})^{j}.
\end{align}
It follows from the symmetric relations \eqref{AF-16} and \eqref{AF-16-1} that $\mathrm{Ker}(\Phi_{\pm}(n,-z_{i}))$ and $\mathrm{Ker}(\Phi_{\pm}(n,\pm z_{i}^{\ast}))$ can be expressed.

We have the evolution of jump matrices
\begin{equation}\nonumber
\begin{aligned}
&G_{1}(t,z)=\textrm{e}^{\delta(z)\omega\hat{\sigma}_{3}t}G_{1}(z),\notag\\
&G_{2,\pm}(t,z)=\textrm{e}^{\delta(z)\omega_{-}\sigma_{3}t}G_{2,\pm}(z)\textrm{e}^{-\delta(z)\omega_{+}\sigma_{3}t},
\end{aligned}
\end{equation}
where $\delta(z)$ and $\omega_{\pm}=\omega(z_{\pm})$ are given by \eqref{AF-27}.

Then we know that the evolution of kernel at $z_{i}$ $(i=1,2,\ldots,k)$ is
\begin{equation}\nonumber
\begin{aligned}
\left(
         \begin{array}{c}
           \omega_{0}(t,z_{i}) \\
           \omega_{1}(t,z_{i}) \\
           \vdots \\
           \omega_{m_{i}-1}(t,z_{i}) \\
         \end{array}
       \right)=&\left(
\begin{array}{cccc}
\textrm{e}^{\delta(z_{i})\omega(z_{i})\sigma_{3}t} & 0 & \cdots & 0 \\
\frac{d\textrm{e}^{\delta(z)\omega(z)\sigma_{3}t}}{dz}|_{z=z_{i}} & \textrm{e}^{\delta(z_{i})\omega(z_{i})\sigma_{3}t} & \cdots & 0 \\
\vdots & \vdots & \ddots & \vdots \\
\frac{d^{m_{i}-1}\textrm{e}^{\delta(z)\omega(z)\sigma_{3}t}}{dz^{m_{i}-1}}|_{z=z_{i}} & \frac{d^{m_{i}-2}\textrm{e}^{\delta(z)\omega(z)\sigma_{3}t}}{dz^{m_{i}-2}}|_{z=z_{i}} & \cdots & \textrm{e}^{\delta(z_{i})\omega(z_{i})\sigma_{3}t} \\
\end{array}
\right)\notag\\
&\times\left(
         \begin{array}{c}
           \omega_{0}(z_{i}) \\
           \omega_{1}(z_{i}) \\
           \vdots \\
           \omega_{m_{i}-1}(z_{i}) \\
         \end{array}
       \right).
\end{aligned}
\end{equation}

\subsection{Riemann-Hilbert problem 2}
In light of the above process, that is the traditional IST method, we cannot obtain high-order rogue waves. Therefore, the robust IST is used to solve this problem. Firstly, a new analytic function needs to be introduced to replace the Jost functions $J_{\pm}(n,z)$. Then we analyze its properties.
\vspace{0.2cm}

\noindent
\textbf{Proposition 2.7}
Let $v_{n}(t)$ satisfy Eq. \eqref{AF-3}, and be a bounded typical solution. Given any integer $n$ and real number $t$. For $z\in M$ $(M = \{z : R^{-1} < |z|\leq R, R > r + A\})$,
$\tau(n, t,z)$ is unique and analytic solution of the Lax pair \eqref{AF-4} under the initial condition
$\tau(0,0,z)=\mathbb{I}_{2}$.
\vspace{0.2cm}

\begin{proof}
The proof is similar to the reference \cite{Y-2021}.
\end{proof}

Considering Proposition 2.7 as well as the existence and uniqueness theorem of ordinary differential equation, we have
\begin{equation}\nonumber
\begin{aligned}
\tau^{in}(n, t,z)=\tau_{o}(n, t,z)\tau_{o}(0,0,z)^{-1}=T^{\pm}(n,t,z)r^{n}\zeta^{n\sigma_{3}}\mu^{\pm}(t,z)T^{\pm}(0,0,z)^{-1}
\end{aligned}
\end{equation}
on the region $M$, in which the matrix function $\tau_{o}(n, t,z)$ is the elementary solution of \eqref{AF-4}.

Moreover, the new sectional analytic matrix function $N(n,t,z)$ reads
\begin{align}\nonumber
N(n,t,z)=\left\{\begin{array}{cc}
                        \tau^{in}(n, t,z)r^{-n}\zeta^{-n\sigma_{3}}\mu^{\pm}(t,z)^{-1}, & z\in M, \\
                        \left(
                                  \begin{array}{cc}
                                    J_{-,1}(n,t,z)a(z)^{-1} & J_{+,2}(n,t,z) \\
                                  \end{array}
                                \right),
                & z\in M_{+}, \\
               \left(
                                  \begin{array}{cc}
                                    J_{+,1}(n,t,z) & J_{-,2}(n,t,z)\bar{a}(z^{\ast})^{-1} \\
                                  \end{array}
                                \right),
               & z\in M_{-}.
                      \end{array}
                      \right.
\end{align}
In the following, we define RHP 2.
\vspace{0.2cm}

\noindent
\textbf{Definition 2.8}
The $2\times 2$ matrix function $N(n,t,z)$ satisfies that
\begin{itemize}
  \item it is analytic in $\mathbb{C}\setminus\left\{\partial M\cup\Sigma\right\}$;
  \item the jump conditions
\begin{equation}\nonumber
\begin{aligned}
N^{+}(n,t,z)=N^{-}(n,t,z)G(n,t,z),~~z\in\partial M\cup\Sigma,
\end{aligned}
\end{equation}
with
\begin{equation}\nonumber
\begin{aligned}
&G(n,t,z)=\zeta^{-2n\sigma_{3}}\textrm{e}^{-2\delta(z)D_{+}\sigma_{3}t},~~~~~z\in\Sigma,\notag\\
&G(n,t,z)=\zeta^{n\sigma_{3}}\textrm{e}^{\delta(z)D\sigma_{3}t}N^{+}(0,0,z)\zeta^{-n\sigma_{3}}
\textrm{e}^{-\delta(z)\omega\sigma_{3}t},~~~~~z\in\partial M_{+},\notag\\
&G(n,t,z)=\zeta^{n\sigma_{3}}\textrm{e}^{\delta(z)D\sigma_{3}t}N^{-}(0,0,\lambda)^{-1}\zeta^{-n\sigma_{3}}
\textrm{e}^{-\delta(z)\omega\sigma_{3}t},~~z\in\partial M_{-};
\end{aligned}
\end{equation}
  \item the normalization conditions
\begin{equation}\nonumber
\begin{aligned}
&N(n,t,z)=\left(
                       \begin{array}{cc}
                         \textrm{e}^{\frac{\textrm{i}}{2}B_{+}}\Gamma_{n}^{-}(t) & 0 \\
                         0 & \textrm{e}^{-\frac{\textrm{i}}{2}B_{-}} \\
                       \end{array}
                     \right)+O(z),~~~~~z\rightarrow0,\notag\\
&N(n,t,z)=\left(
                       \begin{array}{cc}
                         \textrm{e}^{\frac{\textrm{i}}{2}B_{-}} & 0 \\
                         0 & \textrm{e}^{-\frac{\textrm{i}}{2}B_{+}}\Gamma_{n}^{+}(t) \\
                       \end{array}
                     \right)+O\left(\frac{1}{z}\right),~~z\rightarrow\infty.
\end{aligned}
\end{equation}
\end{itemize}
\vspace{0.2cm}

Based on the assumption of function $a(z)$ defined by \eqref{AF-18-1}, the radii of annulus $M$ is assigned a value, which aims to contain all zeros of $a(z)$ and $a^{\ast}(z)$.

Similar to \cite{DB-2019}, it is easy to verify that the existence and uniqueness in Definition 2.8 in term of the Schwartz symmetry properties on the contour $\Omega_{0}$ instead of that on the line. In consequence, the potential function is written as
\begin{align}\label{AF-28}
v_{n}=-\lim_{z\rightarrow\infty}z \frac{N_{1,2}(n,t,z)}{N_{2,2}(n,t,z)}.
\end{align}

\section{Darboux transformation}

The loop group approach \cite{CL-2000} is applied to derive the fundamental Darboux matrix. Combining the fundamental Darboux matrix and a proper RHP, on the basis of the robust IST, the potential functions can be defined.

\subsection{Fundamental Darboux transformation}
For the spectral problem of Eq. \eqref{AF-3}, the fundamental Darboux matrix $V_{1}(n, t,z)$ needs to be deduced.
%We construct the fundamental Darboux matrix $V_{1}(n, t,z)$ for the spectral problem of Eq.\eqref{AF-3}.
Suppose that $\Phi(n,t,z)$ is analytic as $z\in\mathbb{C}\setminus\{0,\infty\}$. In view of the loop group construction, we know that $V_{1}(n, t,z)$ is linear fractional transformation of matrix. We conclude that Darboux matrix satisfies
\begin{enumerate}
  \item the premise about Darboux matrix
\begin{align}\nonumber
&V_{1}(n,t,z)=\left(
                     \begin{array}{cc}
                       1 & 0 \\
                       0 & a_{1}(n,t) \\
                     \end{array}
                   \right)\left(\mathbb{I}+\frac{|x_{1}(n,t)\rangle\langle y_{1}(n,t)|}{z-z_{1}^{\ast}}-\frac{\sigma_{3}|x_{1}(n,t)\rangle\langle y_{1}(n,t)|\sigma_{3}}{z+z_{1}^{\ast}}\right);
\end{align}
  \item the symmetry property
\begin{align}\label{AF-29}
X_{n}^{[1]}(t)=z M_{+}+N_{n}^{[1]}(t)+M_{-}z^{-1}\equiv V_{1}(n+1,t,z)X_{n}V_{1}(n,t,z)^{-1},
\end{align}
with
\begin{align}\nonumber
N_{n}^{[1]}=\left(
         \begin{array}{cc}
           0 & v_{n}^{[1]} \\
           -\overline{v_{n}^{[1]}} & 0 \\
         \end{array}
       \right);
\end{align}
  \item the kernel conditions
\begin{equation}\nonumber
\begin{aligned}
&\mathrm{Ker}(V_{1}(n,t,z_{1}))=\Phi(n,t,z_{1})c_{1},\notag\\
&\mathrm{Ker}(V_{1}(n,t,-z_{1}))=\sigma_{3}\Phi(n,t,z_{1})c_{1},
\end{aligned}
\end{equation}
where $c_{1}$ denotes a column vector;
  \item the residue conditions
\begin{equation}\nonumber
\begin{aligned}
&\mathrm{Res}_{z=z_{1}^{\ast}}(V_{1}(n,t,z_{1})\sigma_{2}\bar{\Phi}(n,t,z_{1})\bar{c}_{1})=0,\notag\\
&\mathrm{Res}_{z=-z_{1}^{\ast}}(V_{1}(n,t,z_{1})\sigma_{2}\sigma_{3}\bar{\Phi}(n,t,z_{1})\bar{c}_{1})=0;
\end{aligned}
\end{equation}
  \item the normalization conditions
\begin{equation}\nonumber
\begin{aligned}
&V_{1}(n,t,z)\rightarrow\left(
                                   \begin{array}{cc}
                                     b_{1}(n,t) & 0 \\
                                     0 & \beta_{1} \\
                                   \end{array}
                                 \right)+O(z),~~~~z\rightarrow0,\notag\\
&V_{1}(n,t,z)\rightarrow\left(
                                   \begin{array}{cc}
                                     1 & 0 \\
                                     0 & c_{1}(n,t) \\
                                   \end{array}
                                 \right)+O\left(\frac{1}{z}\right),~~z\rightarrow\infty,
\end{aligned}
\end{equation}
where a constant $\beta_{1}$ which does not depend on $n$ and $t$, functions $b_{1}(n,t)$ and $c_{1}(n,t)$ need to be determined.
\end{enumerate}
Based on above five conditions, the form of the primary Darboux matrix is
\begin{align}\label{AF-30}
V_{1}(n,t,z)=\left(
                     \begin{array}{cc}
                       1 & 0 \\
                       0 & |z_{1}|^{2}(1+|g_{1}|^{2}\beta^{-1})^{-1} \\
                     \end{array}
                   \right)\left(\mathbb{I}-\frac{z_{1}^{\ast}K^{-1}|y_{1}\rangle\langle y_{1}|}{z-z_{1}^{\ast}}+\frac{z_{1}^{\ast}\sigma_{3}K^{-1}|y_{1}\rangle\langle y_{1}|\sigma_{3}}{z+z_{1}^{\ast}}\right),
\end{align}
where
\begin{equation}\nonumber
\begin{aligned}
&K=\left(
     \begin{array}{cc}
       \alpha & 0 \\
       0 & \beta \\
     \end{array}
   \right),~~|y_{1}\rangle=(f_{1},g_{1})^{\mathrm{T}},~~\langle y_{1}|=(|y_{1}\rangle)^{\dagger},\notag\\
&\alpha=\frac{\langle y_{1}||y_{1}\rangle}{|z_{1}|^{2}-1}-\frac{\langle y_{1}|\sigma_{3}|y_{1}\rangle}{|z_{1}|^{2}+1},~~\beta=\frac{\langle y_{1}||y_{1}\rangle}{|z_{1}|^{2}-1}+\frac{\langle y_{1}|\sigma_{3}|y_{1}\rangle}{|z_{1}|^{2}+1}.
\end{aligned}
\end{equation}
Since $V_{1}(n,t,z)$ is only dependent on $\mathrm{span}(|y_{1}\rangle)$, it stays the same under any rescaling
\begin{align}\nonumber
|y_{1}\rangle\mapsto k|y_{1}\rangle,~~k\in \mathbb{C}.
\end{align}
Eqs. \eqref{AF-29} and \eqref{AF-30} are expanded in the neighborhood of $\infty$, we deduce the corresponding BT as
\begin{align}\label{AF-31}
v_{n}^{[1]}=\frac{(|f_{1}|^{2}+|z_{1}|^{2}|g_{1}|^{2})v_{n}+z_{1}^{\ast}(|z_{1}|^{4}-1)f_{1}\bar{g}_{1}}
{|z_{1}|^{2}|f_{1}|^{2}+|g_{1}|^{2}}.
\end{align}
Then the form of the new analytic matrix solution is
\begin{align}\label{AF-32}
\Phi_{[1]}(n,t,z)=V_{1}(n,t,z)\Phi(n,t,z)V_{1}(0,0,z)^{-1},
\end{align}
and it satisfies \eqref{AF-4}.
%with the new potential function \eqref{AF-31}.

In order to express a determinant form for the solution, the $N$-fold Darboux matrices are represented as follows.
\vspace{0.2cm}

\noindent
\textbf{Proposition 3.1}
If the Lax pair \eqref{AF-4} possesses $N$ different solutions $|y_{i}\rangle$ at $z=z_{i}$,
then we have the Darboux matrix with the form
\begin{align}\label{AF-33}
V_{N}(n,t,z)=\left(
                     \begin{array}{cc}
                       1 & 0 \\
                       0 & a_{N}(n,t) \\
                     \end{array}
                   \right)\left[\mathbb{I}+\sum_{i=1}^{N}\left(\frac{|x_{i}\rangle\langle y_{i}|}{z-z_{i}^{\ast}}-\frac{\sigma_{3}|x_{i}\rangle\langle y_{i}|\sigma_{3}}{z+z_{i}^{\ast}}\right)\right],
\end{align}
for the analytic matrix solution $\Phi(n,t,z)$.
Here
%$|x_{i}\rangle$ is defined by
\begin{equation}\nonumber
\begin{aligned}
&a_{N}(n,t)=\frac{\prod_{i=1}^{N}|z_{i}|^{2}}{1+2Y_{2}\beta^{-1}\Lambda Y_{2}^{\dagger}},~~\Lambda=\left(
     \begin{array}{cccc}
       \bar{z}_{1} & 0 & \cdots & 0 \\
       0 & \bar{z}_{2} & \cdots & 0 \\
       \vdots & \vdots & \ddots & \vdots \\
       0 & 0 & \cdots & \bar{z}_{N} \\
     \end{array}
   \right),\notag\\
&\alpha=\left(\frac{\langle y_{i}||y_{j}\rangle}{z_{j}-z^{\ast}_{i}}-\frac{\langle y_{i}|\sigma_{3}|y_{j}\rangle}{z_{j}+z^{\ast}_{i}}\right)_{1\leq i,j\leq N},~~\beta=\left(\frac{\langle y_{i}||y_{j}\rangle}{z_{j}-z^{\ast}_{i}}+\frac{\langle y_{i}|\sigma_{3}|y_{j}\rangle}{z_{j}+z^{\ast}_{i}}\right)_{1\leq i,j\leq N},\notag\\
&\langle x_{i}|=|x_{i}\rangle^{\dagger},~~\langle y_{i}|=|y_{i}\rangle^{\dagger},~~X_{1}=-Y_{1}\alpha^{-1},~~X_{2}=-Y_{2}\beta^{-1},\notag\\
&\left(
   \begin{array}{c}
     X_{1} \\
     X_{2} \\
   \end{array}
 \right)=\left(
           \begin{array}{cccc}
             |x_{1}\rangle & |x_{2}\rangle & \ldots & |x_{N}\rangle \\
           \end{array}
         \right),~~\left(
   \begin{array}{c}
     Y_{1} \\
     Y_{2} \\
   \end{array}
 \right)=\left(
           \begin{array}{cccc}
             |y_{1}\rangle & |y_{2}\rangle & \ldots & |y_{N}\rangle \\
           \end{array}
         \right).
\end{aligned}
\end{equation}
\vspace{0.2cm}

According to the Darboux matrix $V_{N}(n,t,z)$ in \eqref{AF-33}, the BT between old potential functions and new ones yields
\begin{align}\nonumber
v_{n}^{[N]}=\frac{v_{n}-b_{N}}{a_{N}},
\end{align}
with
\begin{align}\nonumber
b_{N}=2\sum_{i=1}^{N}(|x_{i}\rangle\langle y_{i}|)_{1,2}=-2Y_{1}\alpha^{-1}Y_{1}^{\dagger}.
\end{align}
Therefore, we calculate
\begin{align}\label{AF-34}
v_{n}^{[N]}=\frac{(v_{n}+2Y_{1}\alpha^{-1}Y_{1}^{\dagger})(1+2Y_{2}\beta^{-1}\Lambda Y_{2}^{\dagger})}
{\prod_{i=1}^{N}|z_{i}|^{2}}.
\end{align}
\vspace{0.2cm}

\noindent
\textbf{Proposition 3.2}
When $v_{n} = A$, the general soliton solution $v_{n}^{[N]}$ in \eqref{AF-34} is written as
\begin{align}\label{AF-35}
v_{n}^{[N]}=A\frac{\det(H)}{\det(T)},
\end{align}
where
\begin{equation}\nonumber
\begin{aligned}
T=\left(\frac{\bar{z}_{i}z_{j}\bar{f}_{i}f_{j}+\bar{g}_{i}g_{j}}
{\bar{z}_{i}^{2}z_{j}^{2}-1}\right)_{1\leq i,j\leq N},~~H=\left(\frac{\bar{f}_{i}f_{j}+\bar{z}_{i}z_{j}\bar{g}_{i}g_{j}}
{\bar{z}_{i}^{2}z_{j}^{2}-1}+\frac{\bar{g_{i}}f_{j}}{A\bar{z}_{i}}\right)_{1\leq i,j\leq N}.
\end{aligned}
\end{equation}

\subsection{High-order Darboux matrix}

It is noted that the spectral parameters $z_{i}$ in Proposition 3.2 are different. If we choose the same spectral parameter, is the conclusion valid? In fact, this case is in perfect agreement with the generalized DT. As presented in literature, the generalized DT is derived by the fundamental DT.
We know that the elementary solution at $z=z_{1}$ plays an important role in generating the general Darboux matrix.
%The key to generate the generalized Darboux matrix is to yield the elementary solution at $z=z_{1}$.
Nevertheless, we can not apply it to the Darboux matrix directly. Thus we consider Eq.\eqref{AF-32} in the process of solving high-order Darboux matrix.
\vspace{0.2cm}

\noindent
\textbf{Proposition 3.3}
When $z=z_{1}$, we have
\begin{equation}\nonumber
\begin{aligned}
&|y_{i}^{[j]}\rangle=\Phi^{[j]}(n,t,z_{1})c_{1},\notag\\
&\Phi(n,t,z)=\sum_{j=1}^{\infty}\Phi^{[j]}(n,t,z_{1})(z-z_{1})^{j}.
\end{aligned}
\end{equation}
Then we obtain the high-order Darboux matrix with the form
\begin{align}\nonumber
V_{N}(n,t,z)=\left(
                     \begin{array}{cc}
                       1 & 0 \\
                       0 & a_{N}(n,t) \\
                     \end{array}
                   \right)\left[\mathbb{I}+XL(z,\bar{z}_{1})Y^{\dagger}
                   +\sigma_{3}XL(-z,\bar{z}_{1})Y^{\dagger}\sigma_{3}\right],
\end{align}
here
\begin{equation}\nonumber
\begin{aligned}
&X_{1}=-Y_{1}\alpha^{-1},~~X_{2}=-Y_{2}\beta^{-1},
~~a_{N}(n,t)=\frac{|z_{1}|^{2N}}{1+2\bar{z}_{1}Y_{2}\beta^{-1}Y_{2}^{\dagger}},\notag\\
&X\equiv\left(
   \begin{array}{c}
     X_{1} \\
     X_{2} \\
   \end{array}
 \right)=\left(
           \begin{array}{cccc}
             |x_{1}^{[0]}\rangle & |x_{1}^{[1]}\rangle & \ldots & |x_{1}^{[N-1]}\rangle \\
           \end{array}
         \right),\notag\\
&Y\equiv\left(
   \begin{array}{c}
     Y_{1} \\
     Y_{2} \\
   \end{array}
 \right)=\left(
           \begin{array}{cccc}
             |y_{1}^{[0]}\rangle & |y_{1}^{[1]}\rangle & \ldots & |y_{1}^{[N-1]}\rangle \\
           \end{array}
         \right),\notag\\
&L(z,\bar{z}_{1})=\frac{\bar{z}_{1}\mathbb{I}}{z_{1}\bar{z}_{1}-1}+
\sum_{i=1}^{N-1}\frac{1}{i!}\frac{d^{i}}{dk^{i}}\left(\frac{k}{kz-1}\right)
|_{k=\bar{z}_{1}}(L_{0})^{i},\notag\\
&L_{0}=\left(\delta_{i,j-1}\right)_{1\leq i,j\leq N},
~~\delta_{i,j-1}=\left\{\begin{array}{cc}
1, & i=j-1, \\
0, & i\neq j-1,
\end{array}
\right.\notag\\
&\alpha=(J_{1}^{\dagger}\mu_{-}J_{1}+K_{1}^{\dagger}\mu_{-}K_{1})-(J_{1}^{\dagger}\mu_{+}J_{1}
+K_{1}^{\dagger}\mu_{+}K_{1}),\notag\\
&\beta=(J_{1}^{\dagger}\mu_{-}J_{1}+K_{1}^{\dagger}\mu_{-}K_{1})+(J_{1}^{\dagger}\mu_{+}J_{1}
+K_{1}^{\dagger}\mu_{+}K_{1}),\notag\\
&\mu_{\pm}=\left(\frac{1}{(i-1)!(j-1)!}\frac{d^{i+j-2}}{dk^{i-1}dy^{j-1}}\left(\frac{k}{ky\pm1}\right)
|_{k=\bar{z}_{1},y=z_{1}}\right)_{1\leq i,j\leq N},\notag\\
&J_{1}=f_{1}^{[0]}\mathbb{I}_{N}+\sum_{i=1}^{N-1}f_{1}^{[j]}E^{j},~~E=\left(\delta_{i,j+1}\right)_{1\leq i,j\leq N},\notag\\
&K_{1}=g_{1}^{[0]}\mathbb{I}_{N}+\sum_{i=1}^{N-1}g_{1}^{[j]}E^{j},
~~|y_{1}^{[j]}\rangle=\left(f_{1}^{[j]},g_{1}^{[j]}\right)^{\mathrm{T}}.
\end{aligned}
\end{equation}
\vspace{0.2cm}

We conclude the BT between new potential functions and old ones of the form
\begin{align}\nonumber
v_{n}^{[N]}=\frac{(v_{n}+2Y_{1}\alpha^{-1}Y_{1}^{\dagger})(1+2\bar{z}_{1}Y_{2}\beta^{-1}Y_{2}^{\dagger})}
{|z_{1}|^{2N}}.
\end{align}
Without loss of generality, we extend a high-order pole to several different high-order ones. In a similar way, the expression of high-order solutions is given by next subsection.

\subsection{Riemann-Hilbert problem 3}
Let us reconsider the Darboux matrix under the robust IST. When $R > \max |z_{i}|$, the sectional analytic matrix function is given by
\begin{equation}\nonumber
\begin{aligned}
Q(n,t,z)=\left\{\begin{array}{c}
                        V_{N}(n,t,z),~~~~~~~~~~~~~~~~~~~~~~~~~~~~~~z\in\{z:|z|>R\}\cup\{z:|z|<R^{-1}\}, \\
                        V_{N}(n,t,z)\Phi(n,t,z)V_{N}^{-1}(0,0,z)\Phi^{-1}(n,t,z),~~ z\in\{z:R^{-1}<|z|<R\}.
                      \end{array}
\right.
\end{aligned}
\end{equation}
We next define the RHP 3.
\vspace{0.2cm}

\noindent
\textbf{Definition 3.4}
The $2\times2$ matrix function $Q(n,t,\lambda)$ satisfies that
\begin{itemize}
  \item it is analytic as $z\in\mathbb{C}\setminus\{z: |z|=R^{-1},|z|=R\}$;
  \item the jump relation
\begin{equation}\nonumber
\begin{aligned}
Q^{+}(n,t,z)=Q^{-}(n,t,z)G(n,t,z),~~z\in\{|z|=R^{-1},|z|=R\},
\end{aligned}
\end{equation}
with
\begin{equation}\nonumber
\begin{aligned}
G(n,t,z)=\Phi(n,t,z)G_{N}(0,0,z)\Phi^{-1}(n,t,z);
\end{aligned}
\end{equation}
  \item the normalization conditions
\begin{equation}\nonumber
\begin{aligned}
&Q(n,t,z)\rightarrow\left(
                       \begin{array}{cc}
                         1 & 0 \\
                         0 & a_{N}(n,t) \\
                       \end{array}
                     \right),~~z\rightarrow\infty,\notag\\
&Q(n,t,z)\rightarrow\prod_{i=1}^{N}|z_{i}|^{2}\left(
                       \begin{array}{cc}
                         a_{N}^{-1}(n,t) & 0 \\
                         0 & 1 \\
                       \end{array}
                     \right),~~z\rightarrow0.
\end{aligned}
\end{equation}
\end{itemize}
\vspace{0.2cm}

Under the Definition 3.4, it follows from the recovering formula \eqref{AF-28} that we obtain the potential function
\begin{align}\label{AF-36}
v_{n}^{[N]}(t)=\lim_{z\rightarrow\infty} \frac{A-z Q_{1,2}(n,t,z)}{Q_{2,2}(n,t,z)}.
\end{align}
Based on the above analysis, the jump matrix $G(n,t,z)$ is closely related to the Darboux matrix when $x=t=0$ and an elementary matrix solution of $\Phi(n,t,z)$ possessing a trivial seed solution $v_{n}(t)$. By reference \cite{DB-2020}, the high-order rogue wave and soliton solutions for integrable equations are analyzed effectively in terms of the RHP 3 of the Darboux matrix.

\section{Rational solutions}
In view of the expressions of BT \eqref{AF-31} or \eqref{AF-36}, various rational solutions can be constructed. Firstly, we need to deal with the linear system with $v_{n} = A$ and $z=z_{i}$.

When $z_{i}\neq r\pm A$, and $z_{i}\neq-r\pm A$, we interpolate the seed solution $v_{n} = A$ into the Lax pair \eqref{AF-4} to obtain
\begin{align}\label{AF-37}
\psi_{n+1}=\tau_{i}\psi_{n},~~\psi_{n,t}=(\delta_{i}\tau_{i}+D_{i}\mathbb{I})\psi_{n},
\end{align}
with
\begin{equation}\nonumber
\begin{aligned}
\tau_{i}=\left(
          \begin{array}{cc}
            z_{i} & A \\
            -A & z_{i}^{-1} \\
          \end{array}
        \right),~~\delta_{i}=\delta(z_{i}),~~D_{i}=D(z_{i}),
\end{aligned}
\end{equation}
where $\delta(z)$ and $D(z)$ are given by \eqref{AF-27} and \eqref{AF-27-1}, respectively.

Then $\tau_{i}$ can be diagonally reduced to
\begin{align}\nonumber
\tau_{i}=rG_{i}\zeta_{i}^{\sigma_{3}}G_{i}^{-1},
\end{align}
where
\begin{equation}\nonumber
\begin{aligned}
G_{i}=\left(
     \begin{array}{cc}
       1 & \xi_{i} \\
       \xi_{i} & 1 \\
     \end{array}
   \right),
~~\zeta_{i}=\zeta(z_{i}),~~\xi_{i}=\xi(z_{i}).
\end{aligned}
\end{equation}
Further on, we solve simultaneously the fundamental solution of linear system \eqref{AF-37}, that is,
\begin{align}\label{AF-38}
\psi_{n}(t,z_{i})=r^{n}\textrm{e}^{(a\sin B+b\cos B)A^{2}t}G_{i}\zeta_{i}^{n\sigma_{3}}\textrm{e}^{\delta_{i}\omega_{i}t\sigma_{3}},
~~\omega_{i}=\omega(z_{i}),
\end{align}
where $\omega(z)$ is defined by \eqref{AF-27}.

Combining formulas \eqref{AF-35} with the special vector solutions produces
\begin{align}\label{AF-39}
\left(
  \begin{array}{c}
    f_{i} \\
    g_{i} \\
  \end{array}
\right)=
\frac{\psi_{n}(t,z_{i})}{r^{n}\textrm{e}^{(a\sin B+b\cos B)A^{2}t}}\left(
                                                                          \begin{array}{c}
                                                                            ~~\frac{1}{2}\textrm{e}^{c_{i}+\gamma_{i}} \\
                                                                            -\frac{1}{2}\textrm{e}^{c_{i}-\gamma_{i}} \\
                                                                          \end{array}
                                                                        \right)=\left(
                                                                                  \begin{array}{c}
                                                                                    ~~\sinh(\kappa_{i}+\gamma_{i}) \\
                                                                                    -\sinh(\kappa_{i}-\gamma_{i}) \\
                                                                                  \end{array}
                                                                                \right),
\end{align}
where
\begin{equation}\label{AF-39-1}
\begin{split}
&\gamma_{i}=\frac{\textrm{i}}{2}\arccos\left(\frac{z_{i}-z_{i}^{-1}}{2A}\right),
~~\kappa_{i}=n\ln\zeta_{i}+\delta_{i}\omega_{i}t+c_{i},\\
&\textrm{e}^{-2\gamma_{i}}=-\xi_{i}=\frac{z_{i}-z_{i}^{-1}}{2A}-\textrm{i}\sqrt
{1-\left(\frac{z_{i}-z_{i}^{-1}}{2A}\right)^{2}},
\end{split}
\end{equation}
and $c_{i}$ is a complex constant.

\subsection{Single soliton solution}

When $z_{i}= r\pm A$, and $z_{i}= -r\pm A$, the normalization approach is applied for getting the elementary solution of \eqref{AF-37} in the subsection.
\vspace{0.2cm}

\noindent
\textbf{Theorem 4.1}
The compact formula of single soliton solution is expressed as
\begin{equation}\label{AF-42}
\begin{split}
v_{n}^{[1]}=&A\left[\cosh(\kappa_{1}+\bar{\kappa}_{1}+B_{1}+2(\gamma_{1}+\bar{\gamma}_{1}))
-r_{1}^{-1}r_{2}\cosh(\kappa_{1}-\bar{\kappa}_{1}+B_{2}+2(\gamma_{1}-\bar{\gamma}_{1}))\right]\\
&\times\left[\cosh(\kappa_{1}+\bar{\kappa}_{1}+B_{1})
-r_{1}^{-1}r_{2}\cosh(\kappa_{1}-\bar{\kappa}_{1}+B_{2})\right]^{-1}.
\end{split}
\end{equation}
\vspace{0.2cm}

\begin{proof}
Inserting \eqref{AF-39} into Eq. \eqref{AF-31} yields
\begin{align}\label{AF-40}
v_{n}^{[1]}=\frac{(|f_{1}|^{2}+|z_{1}|^{2}|g_{1}|^{2})A+\bar{z}_{1}^{-1}(|z_{1}|^{4}-1)\bar{g}_{1}f_{1}}
{|z_{1}|^{2}|f_{1}|^{2}+|g_{1}|^{2}}.
\end{align}
In order to obtain the properties of soliton solution, we simplify \eqref{AF-40} and calculate
\begin{equation}\nonumber
\begin{aligned}
&|z_{1}|^{2}|f_{1}|^{2}+|g_{1}|^{2}=r_{1}\cosh(\kappa_{1}+\bar{\kappa}_{1}+B_{1})
-r_{2}\cosh(\kappa_{1}-\bar{\kappa}_{1}+B_{2}),\notag\\
&|f_{1}|^{2}+|z_{1}|^{2}|g_{1}|^{2}+(A\bar{z}_{1})^{-1}
(|z_{1}|^{4}-1)\bar{g}_{1}f_{1}\notag\\
&~~~~~~~~~~~~~~~~~~~~~=r_{3}\cosh(\kappa_{1}+\bar{\kappa}_{1}+B_{3})
-r_{4}\cosh(\kappa_{1}-\bar{\kappa}_{1}+B_{4}),
\end{aligned}
\end{equation}
where
\begin{equation}\nonumber
\begin{aligned}
&B_{i}=\frac{1}{2}\ln\left(p_{i}(q_{i})^{-1}\right),~~r_{i}=\sqrt{p_{i}q_{i}},~~i=1,2,3,4,\notag\\
&p_{1}=|z_{1}|^{2}\textrm{e}^{(\gamma_{1}-\bar{\gamma}_{1})}+\textrm{e}^{-(\gamma_{1}-\bar{\gamma}_{1})},
~~q_{1}=|z_{1}|^{2}\textrm{e}^{-(\gamma_{1}-\bar{\gamma}_{1})}+\textrm{e}^{(\gamma_{1}-\bar{\gamma}_{1})},\notag\\
&p_{2}=|z_{1}|^{2}\textrm{e}^{(\gamma_{1}+\bar{\gamma}_{1})}+\textrm{e}^{-(\gamma_{1}+\bar{\gamma}_{1})},
~~q_{2}=|z_{1}|^{2}\textrm{e}^{-(\gamma_{1}+\bar{\gamma}_{1})}+\textrm{e}^{(\gamma_{1}+\bar{\gamma}_{1})},\notag\\
&p_{3}=-\frac{|z_{1}|^{2}\textrm{e}^{2\gamma_{1}}+\textrm{e}^{2\bar{\gamma}_{1}}}{|z_{1}|^{2}
\textrm{e}^{-2\gamma_{1}}+\textrm{e}^{-i\bar{\gamma}_{1}}}p_{1},
~~~~q_{3}=-\frac{|z_{1}|^{2}\textrm{e}^{-2\gamma_{1}}+\textrm{e}^{-2\bar{\gamma}_{1}}}{|z_{1}|^{2}\textrm{e}^{2\gamma_{1}}
+\textrm{e}^{2\bar{\gamma}_{1}}}q_{1},\notag\\
&p_{4}=-\frac{|z_{1}|^{2}\textrm{e}^{-2\gamma_{1}}+\textrm{e}^{2\bar{\gamma}_{1}}}{|z_{1}|^{2}
\textrm{e}^{2\gamma_{1}}+\textrm{e}^{-2\bar{\gamma}_{1}}}p_{2},
~~~~~~q_{4}=-\frac{|z_{1}|^{2}\textrm{e}^{2\gamma_{1}}+\textrm{e}^{-2\bar{\gamma}_{1}}}{|z_{1}|^{2}
\textrm{e}^{-2\gamma_{1}}+\textrm{e}^{2\bar{\gamma}_{1}}}q_{2}.
\end{aligned}
\end{equation}
It is noted that $r_{3}=r_{1}$, $r_{4}=r_{2}$, $B_{3}=B_{1}+2(\gamma_{1}+\bar{\gamma}_{1})$ and
$B_{4}=B_{2}+2(\gamma_{1}-\bar{\gamma}_{1})$.
Thus, we obtain
\begin{align}\nonumber
v_{n}^{[1]}=\frac{A[r_{3}\cosh(\kappa_{1}+\bar{\kappa}_{1}+B_{3})
-r_{4}\cosh(\kappa_{1}-\bar{\kappa}_{1}+B_{4})]}
{r_{1}\cosh(\kappa_{1}+\bar{\kappa}_{1}+B_{1})
-r_{2}\cosh(\kappa_{1}-\bar{\kappa}_{1}+B_{2})}.
\end{align}
This proves the theorem.
\end{proof}

It follows from \eqref{AF-42} that the velocity of soliton is $-\mathrm{Re}(\delta_{1}\omega_{1})\mathrm{Re}(\ln (\zeta_{1}))^{-1}$. Define the line $\mathcal{L}$ of the form
\begin{align}\nonumber
2\mathrm{Im}(\ln(\zeta_{1}))\left(n+\frac{\mathrm{Im}(\delta_{1}\omega_{1})}{\mathrm{Im}(\ln (\zeta_{1}))}t\right)=\tilde{c},
\end{align}
where $\tilde{c}$ ia a constant. The soliton solutions are going to oscillate along lines perpendicular to $\mathcal{L}$ when they show the breather dynamics.

Based on the expression of Eq. \eqref{AF-42}, when $\kappa_{1}+\bar{\kappa}_{1}+B_{1}\rightarrow\pm\infty$, one arrives at
\begin{equation}\nonumber
\begin{aligned}
v_{n}^{[1]}\rightarrow\left\{\begin{array}{cc}
                        A\textrm{e}^{2(\gamma_{1}+\bar{\gamma}_{1})},~ & \kappa_{1}+\bar{\kappa}_{1}+B_{1}\rightarrow+\infty, \\
                        A\textrm{e}^{-2(\gamma_{1}+\bar{\gamma}_{1})}, & \kappa_{1}+\bar{\kappa}_{1}+B_{1}\rightarrow-\infty.
                      \end{array}
                      \right.
\end{aligned}
\end{equation}
As a result, when $-4\textrm{i}(\gamma_{1}+\bar{\gamma}_{1})\mod(2\pi)\neq 0$, there is a relatively large phase difference. It follows from \eqref{AF-42} that we have the velocity and phase difference of the localized wave solution. Moreover, we introduce the following theorem to calculate the maximum value.
\vspace{0.2cm}

\noindent
\textbf{Theorem 4.2}
Given $|z_{1}|>1$, suppose that
\begin{equation}\nonumber
 \begin{split}
c_{1}=\frac{1}{2}\ln\left(\frac{1+(r+A)z_{1}\textrm{e}^{-2\gamma_{1}}}
{(r+A)z_{1}+\textrm{e}^{-2\gamma_{1}}}\right),
 \end{split}
\end{equation}
we conclude that the maximal modulus of solution \eqref{AF-40} is
%that the modulus of solution \eqref{AF-40} achieves the maximum
$\frac{1}{2}[(r+A)|z_{1}|^{2}-(r-A)|z_{1}|^{-2}]$ when $n=t=0$.
\vspace{0.2cm}

\begin{proof}
Let's first recall Eq. \eqref{AF-40}, then rewrite it as
\begin{align}\label{AF-42-1}
v_{n}^{[1]}=\frac{(1+|z_{1}|^{2}|f_{1}^{-1}g_{1}|^{2})A+\bar{z}_{1}^{-1}(|z_{1}|^{4}-1)
\bar{f}_{1}^{-1}\bar{g}_{1}}{|z_{1}|^{2}+|f_{1}^{-1}g_{1}|^{2}}.
\end{align}
To deduce the maximum from \eqref{AF-42-1}, taking
\begin{equation}\nonumber
 \begin{split}
f_{1}=1,~~g_{1}=(r+A)z_{1},
 \end{split}
\end{equation}
we calculate
\begin{equation}\nonumber
 \begin{split}
\max|v_{n}^{[1]}|=\frac{1}{2}[(r+A)|v_{1}|^{2}-(r-A)|v_{1}|^{-2}].
 \end{split}
\end{equation}
It is obvious that when
\begin{equation}\nonumber
 \begin{split}
\left(g_{1}(0,0), f_{1}(0,0)\right)=(k(r+A)z_{1}, k),~~k\in \mathbb{C},
 \end{split}
\end{equation}
$|v_{n}^{[1]}|$ arrives at the maximum value with $(n,t)=(0,0)$. In fact, we can choose
\begin{equation}\nonumber
 \begin{split}
c_{1}=\frac{1}{2}\ln\left(\frac{1+(r+A)z_{1}\textrm{e}^{-2\gamma_{1}}}{(r+A)z_{1}
+\textrm{e}^{-2\gamma_{1}}}\right)
 \end{split}
\end{equation}
to obtain the maximum.
\end{proof}

According to the expression \eqref{AF-42} of single soliton solution, we exhibit the dynamic behaviors and wave propagation patterns.

\begin{figure}
\begin{center}
{\rotatebox{0}{\includegraphics[width=5.8cm,height=4.0cm,angle=0]{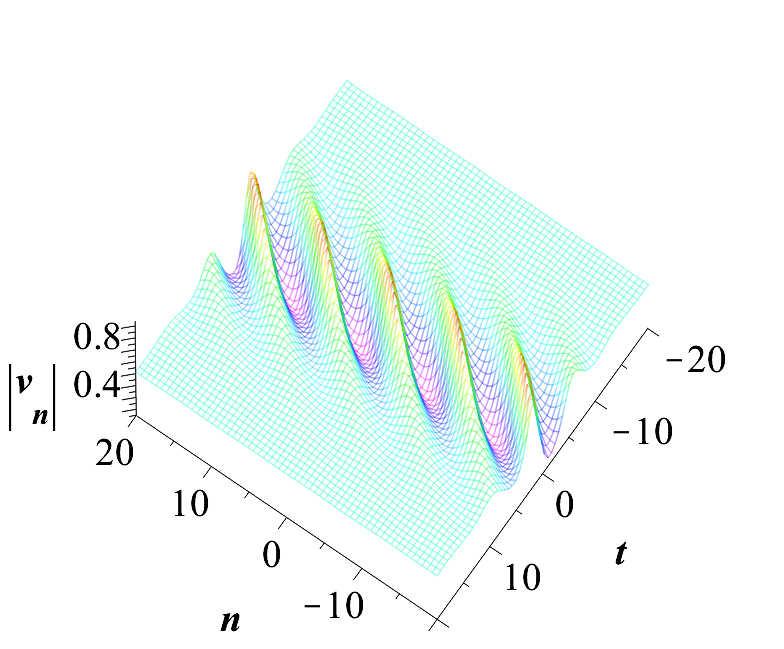}}}
~~{\rotatebox{0}{\includegraphics[width=5.8cm,height=4.0cm,angle=0]{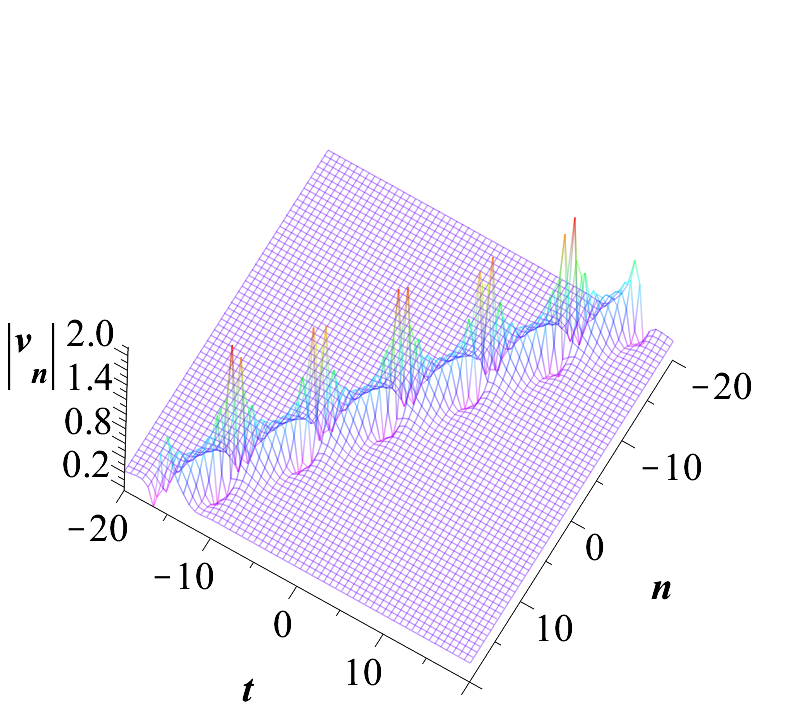}}}\\
$\hspace{0em}\textbf{(a)}
\hspace{18em}\textbf{(b)}$\\
{\rotatebox{0}{\includegraphics[width=5.8cm,height=4.0cm,angle=0]{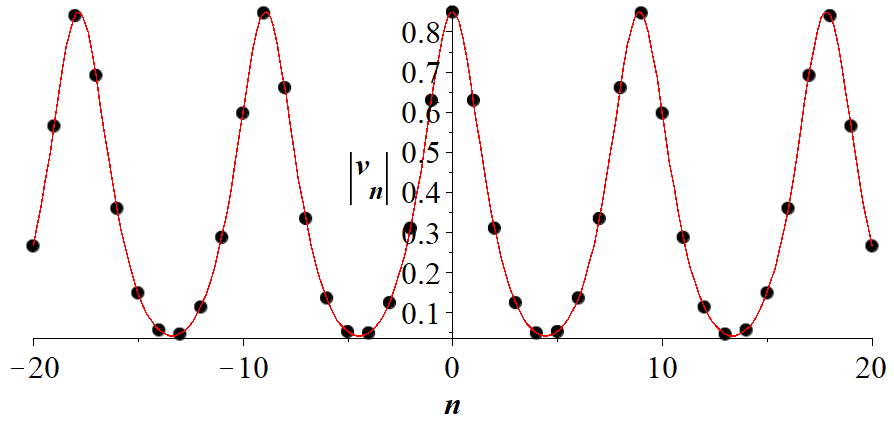}}}
~~{\rotatebox{0}{\includegraphics[width=5.8cm,height=4.0cm,angle=0]{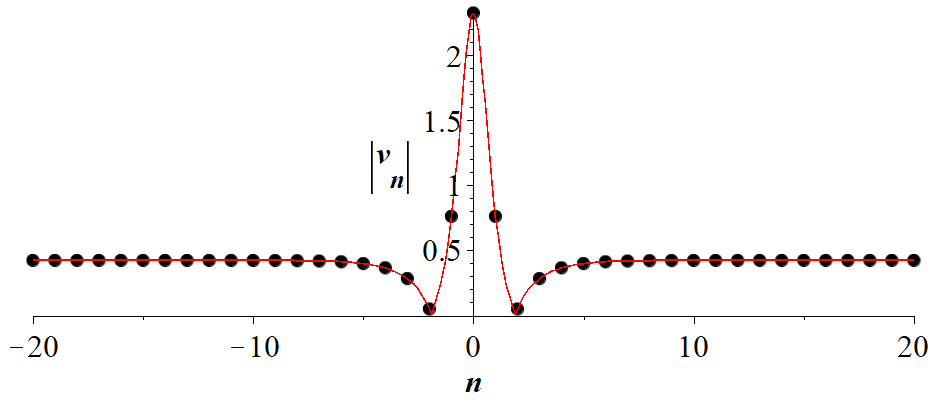}}}\\
$\hspace{0em}\textbf{(c)}
\hspace{18em}\textbf{(d)}$
\end{center}
\end{figure}
{\small Figure 2\quad The single soliton solution in \eqref{AF-42} with $a=1.0, b=0.5, A=\frac{5}{12}, B=0$:
$\textbf{(a)}$ $z_{1}=1.2$;
$\textbf{(b)}$ $z_{1}=1.8$;
wave propagation pattern of the wave along with the $n$ axis:
$\textbf{(c)}$ $\max|v_{n}^{[1]}|\approx0.85$;
$\textbf{(d)}$ $\max|v_{n}^{[1]}|\approx2.33$.}

\begin{figure}
\begin{center}
{\rotatebox{0}{\includegraphics[width=5.8cm,height=4.0cm,angle=0]{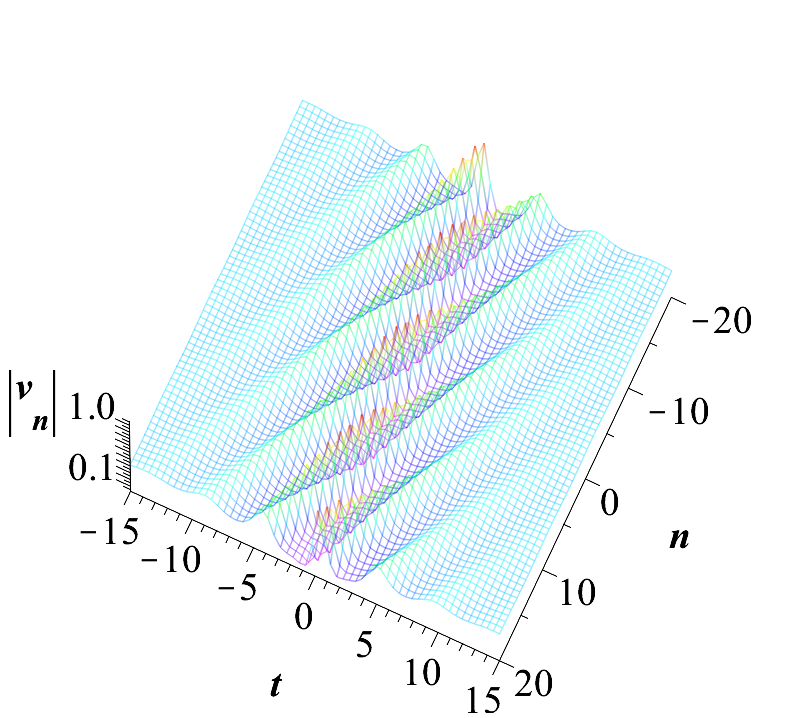}}}
~~{\rotatebox{0}{\includegraphics[width=5.8cm,height=4.0cm,angle=0]{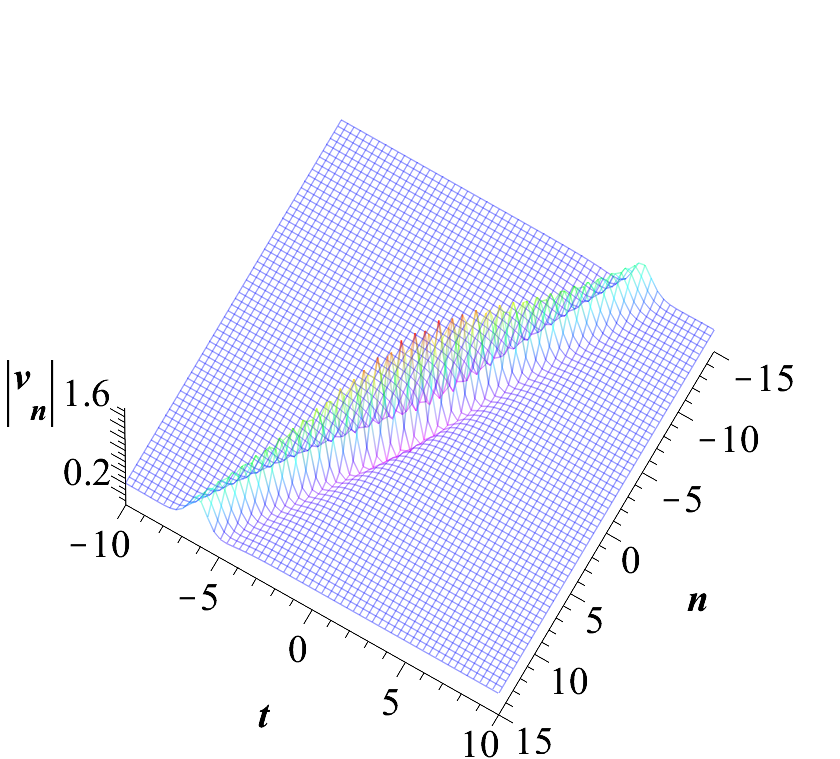}}}\\
$\hspace{0em}\textbf{(a)}
\hspace{18em}\textbf{(b)}$\\
{\rotatebox{0}{\includegraphics[width=5.8cm,height=4.0cm,angle=0]{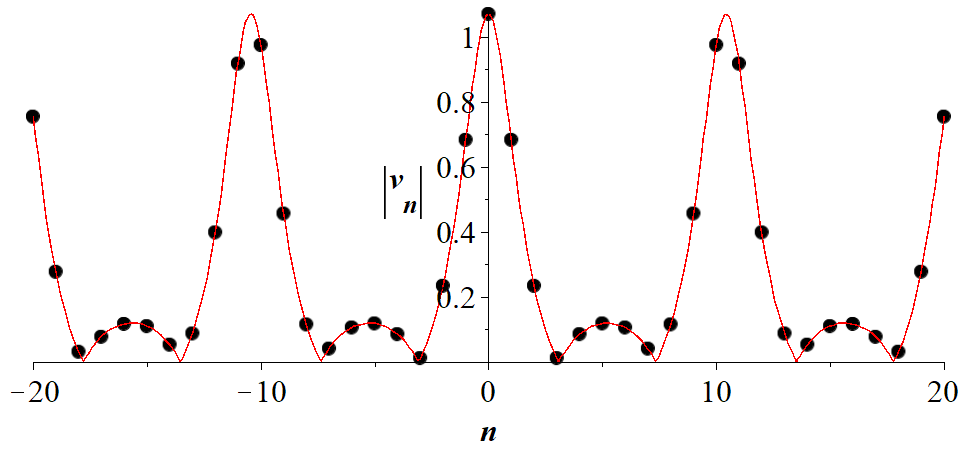}}}
~~{\rotatebox{0}{\includegraphics[width=5.8cm,height=4.0cm,angle=0]{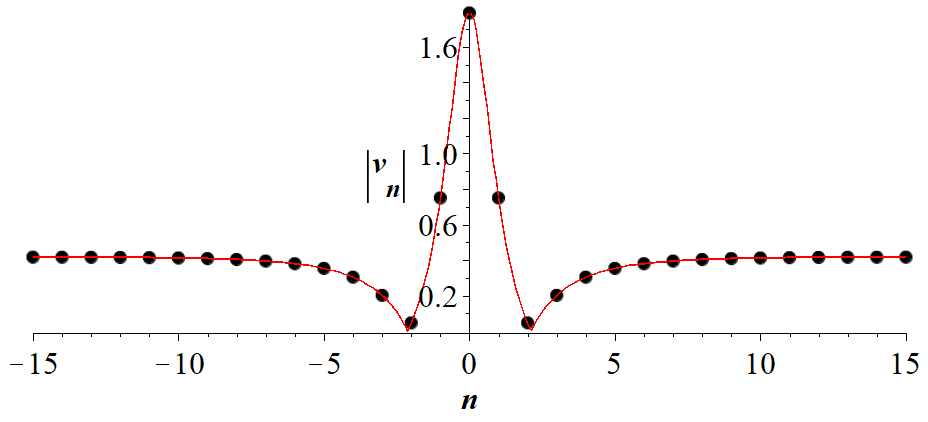}}}\\
$\hspace{0em}\textbf{(c)}
\hspace{18em}\textbf{(d)}$
\end{center}
\end{figure}
{\small Figure 3\quad The single soliton solution in \eqref{AF-42} with $a=1.0, A=\frac{5}{12}, B=\frac{\pi}{2}$:
$\textbf{(a)}$ $ b=1.0, z_{1}=1.3$;
$\textbf{(b)}$ $ b=0.5, z_{1}=1.6$;
wave propagation pattern of the wave along with the $n$ axis:
$\textbf{(c)}$ $\max|v_{n}^{[1]}|\approx1.07$;
$\textbf{(d)}$ $\max|v_{n}^{[1]}|\approx1.79$.}
\vspace{0.2cm}

Under the modulational unstable background, without loss of generality, we take $B=0$ and obtain the Akhmediev breather and the Kuznetsov-Ma breather, as shown in Figs. 2(a) and (b), respectively.
Under the modulational stable background, we take $B=\frac{\pi}{2}$ and acquire the periodic solution and the W-shape soliton, as shown in Figs. 3(a) and (b), respectively.

\subsection{$N$-soliton solution}

We investigate the interaction law for the $N$-soliton solutions with the parameters $z_{i}$, $c_{i}$, $|z_{i}|>1$, and $\mathrm{Re}(\ln(\zeta_{i}))>0$, $i=1,2,\ldots,N$. Then we arrange the velocity parameters with the order $s_{1} < s_{2} < \cdots < s_{N}$, where
\begin{align}\nonumber
s_{i}=-\frac{\mathrm{Re}(\delta_{i}\omega_{i})}{\mathrm{Re}(\ln(\zeta_{i}))}.
\end{align}
The asymptotic behavior of the $k$th localized wave solution is given by the following theorem.
\vspace{0.2cm}

\noindent
\textbf{Theorem 4.3}
The asymptotic behavior along the line $n-s_{k}t = const$ as $t$ tends to plus and minus infinity of $N$-soliton solutions can be written as
\begin{equation}\label{AF-43-1}
\begin{split}
v_{n}^{[N]}=&A\left[\cosh(\kappa_{k}^{\pm}+\bar{\kappa}_{k}^{\pm}+B_{1}^{(k)}
+2(\gamma_{k}+\bar{\gamma}_{k}))\right.\\
&\left.
-r_{2}^{(k)}(r_{1}^{(k)})^{-1}\cosh(\kappa_{k}^{\pm}-\bar{\kappa}_{k}^{\pm}+B_{2}^{(k)}
+2(\gamma_{k}-\bar{\gamma}_{k}))\right]\\
&\times\left[\cosh(\kappa_{k}^{\pm}+\bar{\kappa}_{k}^{\pm}+B_{1}^{(k)})
-r_{2}^{(k)}(r_{1}^{(k)})^{-1}\cosh(\kappa_{k}^{\pm}-\bar{\kappa}_{k}^{\pm}+B_{2}^{(k)})\right]^{-1}\\
&\times\textrm{e}^{\textrm{i}\Gamma_{(k)}^{\pm}}+O\left(\textrm{e}^{-c^{(k)}|t|}\right),
\end{split}
\end{equation}
where
\begin{equation}\nonumber
\begin{aligned}
\kappa_{k}^{\pm}&=n\ln\zeta_{k}+\delta_{k}\omega_{k}t+c_{k}+\frac{1}{2}
\ln\left(\omega_{k}^{\pm}(D_{k}^{\pm})^{-1}\right),\notag\\
B_{i}^{(k)}&=\frac{1}{2}\ln\left(p_{i}^{(k)}(q_{i}^{(k)})^{-1}\right),
~~r_{i}^{(k)}=\sqrt{p_{i}^{(k)}q_{i}^{(k)}},~~i=1,2,\notag\\
p_{1}^{(k)}&=|z_{k}|^{2}\textrm{e}^{(\gamma_{k}-\bar{\gamma}_{k})}
+\textrm{e}^{-(\gamma_{k}-\bar{\gamma}_{k})},
~~q_{1}^{(k)}=|z_{k}|^{2}\textrm{e}^{-(\gamma_{k}-\bar{\gamma}_{k})}
+\textrm{e}^{(\gamma_{k}-\bar{\gamma}_{k})},\notag\\
p_{2}^{(k)}&=|z_{k}|^{2}\textrm{e}^{(\gamma_{k}+\bar{\gamma}_{k})}
+\textrm{e}^{-(\gamma_{k}+\bar{\gamma}_{k})},
~~q_{2}^{(k)}=|z_{k}|^{2}\textrm{e}^{-(\gamma_{k}+\bar{\gamma}_{k})}
+\textrm{e}^{(\gamma_{k}+\bar{\gamma}_{k})}.
\end{aligned}
\end{equation}
\vspace{0.2cm}

\begin{proof}
We first decompose the $N$-fold Darboux matrix into two Darboux matrices
\begin{align}\nonumber
V_{N}(n,t,z)=V^{[k]}(n,t,z)V_{(k)}(n,t,z),
\end{align}
with
\begin{equation}\label{AF-43}
\begin{split}
&V^{[k]}(n,t,z)=\left(
                     \begin{array}{cc}
                       1 & 0 \\
                       0 & \frac{|z_{k}|^{2}}{1+2|\hat{g}_{k}|^{2}\beta_{k}^{-1}} \\
                     \end{array}
                   \right)\left(\mathbb{I}-\frac{z_{k}^{\ast}K^{-1}|\hat{y}_{k}\rangle\langle \hat{y}_{k}|}{z-z_{k}^{\ast}}+\frac{z_{k}^{\ast}\sigma_{3}K^{-1}
                   |\hat{y}_{k}\rangle\langle \hat{y}_{k}|\sigma_{3}}{z+z_{k}^{\ast}}\right),\\
&V_{(k)}(n,t,z)=\left(
                     \begin{array}{cc}
                       1 & 0 \\
                       0 & a_{(k)}(n,t) \\
                     \end{array}
                   \right)\left[\mathbb{I}+\sum_{i=1,i\neq k}^{N}\left(\frac{|x_{i}^{(k)}\rangle\langle y_{i}|}{z-z_{i}^{\ast}}-\frac{\sigma_{3}|x_{i}^{(k)}\rangle\langle y_{i}|\sigma_{3}}{z+z_{i}^{\ast}}\right)\right],
\end{split}
\end{equation}
where
\begin{equation}\nonumber
\begin{aligned}
&X_{1}^{(k)}=-Y_{1}^{(k)}\alpha_{(k)}^{-1},~~X_{2}^{(k)}=-Y_{2}^{(k)}\beta_{(k)}^{-1},\notag\\
&\alpha_{k}=\frac{\langle \hat{y}_{k}||\hat{y}_{k}\rangle}{|z_{k}|^{2}-1}-\frac{\langle \hat{y}_{k}|\sigma_{3}|\hat{y}_{k}\rangle}{|z_{k}|^{2}+1},~~\beta_{k}=\frac{\langle \hat{y}_{k}||\hat{y}_{k}\rangle}{|z_{k}|^{2}-1}+\frac{\langle \hat{y}_{k}|\sigma_{3}|\hat{y}_{k}\rangle}{|z_{k}|^{2}+1},\notag\\
&K=\left(
     \begin{array}{cc}
       \alpha_{k} & 0 \\
       0 & \beta_{k} \\
     \end{array}
   \right),~~|\hat{y}_{k}\rangle=(\hat{f}_{k},\hat{g}_{k})^{\mathrm{T}}
   =V_{(k)}(n,t,z)|y_{k}\rangle,\notag\\
&a_{(k)}(n,t)=\frac{\prod_{i=1,i\neq k}^{N}|z_{i}|^{2}}{1+2Y_{2}^{(k)}\beta_{(k)}^{-1}\Lambda_{(k)}(Y_{2}^{(k)})^{\dagger}},
~~\Lambda_{(k)}=\mathrm{diag}(\bar{z}_{1}, \bar{z}_{2}, \cdots, \bar{z}_{N})
,\notag\\
&\left(
   \begin{array}{c}
     X_{1}^{(k)} \\
     X_{2}^{(k)} \\
   \end{array}
 \right)=\left(
           \begin{array}{cccc}
             |x_{1}^{(k)}\rangle & |x_{2}^{(k)}\rangle & \ldots & |x_{N}^{(k)}\rangle \\
           \end{array}
         \right),~~\left(
   \begin{array}{c}
     Y_{1}^{(k)} \\
     Y_{2}^{(k)} \\
   \end{array}
 \right)=\left(
           \begin{array}{cccc}
             |y_{1}\rangle & |y_{2}\rangle & \ldots & |y_{N}\rangle \\
           \end{array}
         \right),\notag\\
&\alpha_{(k)}=\left(\frac{\langle y_{i}||y_{j}\rangle}{z_{j}-z_{i}^{\ast}}-\frac{\langle y_{i}|\sigma_{3}|y_{j}\rangle}{z_{j}+z_{i}^{\ast}}\right)_{1\leq i,j\leq N,i,j\neq k},~~\beta_{(k)}=\left(\frac{\langle y_{i}||y_{j}\rangle}{z_{j}-z_{i}^{\ast}}+\frac{\langle y_{i}|\sigma_{3}|y_{j}\rangle}{z_{j}+z_{i}^{\ast}}\right)_{1\leq i,j\leq N,i,j\neq k}.
\end{aligned}
\end{equation}
Then, along the line
\begin{equation}\nonumber
\begin{aligned}
&\mathrm{Re}(\ln(\zeta_{k}))(n-s_{k}t) = \tilde{c},\notag\\
&\mathrm{Re}(\kappa_{i}) = \mathrm{Re}(\ln(\zeta_{i}))[n-s_{k}t+(s_{k}-s_{i})t],
\end{aligned}
\end{equation}
and when $t\rightarrow\pm\infty$, we have
\begin{equation}\nonumber
\begin{aligned}
\mathrm{Re}(\kappa_{i})\rightarrow\left\{\begin{array}{cc}
                                   \pm\infty, & i<k, \\
                                   \mp\infty, & i>k.
                                 \end{array}
                                 \right.
\end{aligned}
\end{equation}
Up to a scalar function, as $t$ goes to plus infinity, the asymptotic expressions for $|y_{i}\rangle$ $(i\neq k)$ are
\begin{equation}\nonumber
\begin{aligned}
|y_{i}\rangle\propto|y_{i}^{+}\rangle=\left\{\begin{array}{cc}
                                               \left(
                                         \begin{array}{c}
                                           1 \\
                                           \xi_{1} \\
                                         \end{array}
                                       \right)+O(\textrm{e}^{-c_{i}^{(k)}|t|}), & i<k, \\
                                               \left(
                                         \begin{array}{c}
                                           \xi_{1} \\
                                           1 \\
                                         \end{array}
                                       \right)+O(\textrm{e}^{-c_{i}^{(k)}|t|}), & i>k,
                                             \end{array}
\right.
\end{aligned}
\end{equation}
and as $t$ goes to minus infinity, the asymptotic expressions for $|y_{i}\rangle$ $(i\neq k)$ are
\begin{equation}\nonumber
\begin{aligned}
|y_{i}\rangle\propto|y_{i}^{-}\rangle=\left\{\begin{array}{cc}
                                               \left(
                                         \begin{array}{c}
                                           \xi_{1} \\
                                           1 \\
                                         \end{array}
                                       \right)+O(\textrm{e}^{-c_{i}^{(k)}|t|}), & i<k, \\
                                               \left(
                                         \begin{array}{c}
                                           1 \\
                                           \xi_{1} \\
                                         \end{array}
                                       \right)+O(\textrm{e}^{-c_{i}^{(k)}|t|}), & i>k,
                                             \end{array}
\right.
\end{aligned}
\end{equation}
where
\begin{align}\nonumber
c_{i}^{(k)}=4\mathrm{Re}(\mathrm{ln}(\zeta_{i}))|s_{i}-s_{k}|.
\end{align}
When $t\rightarrow\pm\infty$, $V_{(k)}(n,t,z)$ in \eqref{AF-43} is rewritten as
\begin{equation}\nonumber
\begin{aligned}
V_{(k)}(n,t,z)=V_{(k)}^{\pm}(z)+O(\textrm{e}^{-c^{(k)}|t|}),
\end{aligned}
\end{equation}
where
\begin{equation}\nonumber
\begin{aligned}
&V_{(k)}^{\pm}(z)=\left(
                     \begin{array}{cc}
                       1 & 0 \\
                       0 & a_{(k)}^{\pm} \\
                     \end{array}
                   \right)\left[\mathbb{I}+\sum_{i=1,i\neq k}^{N}\left(\frac{|x_{i}^{(k)\pm}\rangle\langle y_{i}^{\pm}|}{z-z_{i}^{\ast}}-\frac{\sigma_{3}|x_{i}^{(k)\pm}\rangle\langle y_{i}^{\pm}|\sigma_{3}}{z+z_{i}^{\ast}}\right)\right],\notag\\
&c^{(k)}=4\min_{i\neq k} (\mathrm{Re}(\ln(\zeta_{i}))|s_{i}-s_{k}|),
~~a_{(k)}^{\pm}(n,t)=\frac{\prod_{i=1,i\neq k}^{N}|z_{i}|^{2}}{1+2Y_{2}^{(k)\pm}(\beta_{\pm}^{(k)})^{-1}\Lambda_{(k)}(Y_{2}^{(k)\pm})^{\dagger}},\notag\\
&\left(
   \begin{array}{c}
     X_{1}^{(k)\pm} \\
     X_{2}^{(k)\pm} \\
   \end{array}
 \right)=\left(
           \begin{array}{cccc}
             |x_{1}^{(k)\pm}\rangle & |x_{2}^{(k)\pm}\rangle & \ldots & |x_{N}^{(k)\pm}\rangle \\
           \end{array}
         \right),\notag\\
&\left(
   \begin{array}{c}
     Y_{1}^{(k)\pm} \\
     Y_{2}^{(k)\pm} \\
   \end{array}
 \right)=\left(
           \begin{array}{cccc}
             |y_{1}^{\pm}\rangle & |y_{2}^{\pm}\rangle & \ldots & |y_{N}^{\pm}\rangle \\
           \end{array}
         \right),\notag\\
&\alpha_{\pm}^{(k)}=\left(\frac{\langle y_{i}^{\pm}||y_{j}^{\pm}\rangle}{z_{j}-z_{i}^{\ast}}-\frac{\langle y_{i}^{\pm}|\sigma_{3}|y_{j}^{\pm}\rangle}{z_{j}+z_{i}^{\ast}}\right)_{1\leq i,j\leq N,i,j\neq k},\notag\\
&\beta_{\pm}^{(k)}=\left(\frac{\langle y_{i}^{\pm}||y_{j}^{\pm}\rangle}{z_{j}-z_{i}^{\ast}}+\frac{\langle y_{i}^{\pm}|\sigma_{3}|y_{j}^{\pm}\rangle}{z_{j}+z_{i}^{\ast}}\right)_{1\leq i,j\leq N,i,j\neq k},\notag\\
&X_{1}^{(k)\pm}=-Y_{1}^{(k)}(\alpha_{\pm}^{(k)})^{-1},~~X_{2}^{(k)\pm}=
-Y_{2}^{(k)}(\beta_{\pm}^{(k)})^{-1}.
\end{aligned}
\end{equation}
The asymptotic behavior of $|\hat{y}_{k}\rangle$ is
\begin{align}\nonumber
|\hat{y}_{k}\rangle=V_{(k)}(n,t,z)|y_{k}\rangle=V_{(k)}^{\pm}(z)|y_{k}\rangle+O(\textrm{e}^{-c^{(k)}|t|}).
\end{align}
Further on, we compute
%In order to get the asymptotic behavior of the $k$th soliton solution, we need to compute more specifically for $V_{(k)}^{\pm}(z_{k})|y_{k}\rangle$,
\begin{equation}\nonumber
\begin{aligned}
V_{(k)}^{\pm}(z_{k})|y_{k}\rangle&=V_{(k)}^{\pm}(z_{k})\left(
                                                            \begin{array}{cc}
                                                              1 & \xi_{k} \\
                                                              \xi_{k} & 1 \\
                                                            \end{array}
                                                          \right)\left(
                                                                   \begin{array}{c}
                                                                     \textrm{e}^{\kappa_{k}} \\
                                                                     \textrm{e}^{-\kappa_{k}} \\
                                                                   \end{array}
                                                                 \right)\notag\\
&=\left(
    \begin{array}{cc}
      \omega_{k}^{\pm} & D_{(k)}^{\pm}\xi_{(k)}\textrm{e}^{\textrm{i}\Gamma_{(k)}^{\pm}} \\
      \omega_{k}^{\pm}\xi_{(k)}\textrm{e}^{-\textrm{i}\Gamma_{(k)}^{\pm}} & D_{k}^{\pm} \\
    \end{array}
  \right)\left(
                                                                   \begin{array}{c}
                                                                     \textrm{e}^{\kappa_{k}} \\
                                                                     \textrm{e}^{-\kappa_{k}} \\
                                                                   \end{array}
                                                                 \right),
\end{aligned}
\end{equation}
where
$\kappa_{k}$ is defined by \eqref{AF-39-1} when $i=k$,
\begin{equation}\nonumber
\begin{aligned}
&\mu_{1}^{(k)\pm}=\left(\frac{\langle y_{i}^{\pm}||y_{k}^{\pm}\rangle}{z_{k}-z_{i}^{\ast}}-\frac{\langle y_{i}^{\pm}|\sigma_{3}|y_{k}^{\pm}\rangle}{z_{k}+z_{i}^{\ast}}\right)_{1\leq i\leq N,i\neq k},\notag\\
&\mu_{2}^{(k)\pm}=\left(\frac{\langle y_{i}^{\pm}||y_{k}^{\pm}\rangle}{z_{k}-z_{i}^{\ast}}+\frac{\langle y_{i}^{\pm}|\sigma_{3}|y_{k}^{\pm}\rangle}{z_{k}+z_{i}^{\ast}}\right)_{1\leq i\leq N,i\neq k},\notag\\
&\omega_{k}^{\pm}=1-Y_{1}^{(k)\pm}(\alpha_{\pm}^{(k)})^{-1}\mu_{1}^{(k)\pm},D_{(k)}^{\pm}=
a_{(k)}^{\pm}\left(1-Y_{2}^{(k)\pm}(\beta_{\pm}^{(k)})^{-1}\right)\mu_{2}^{(k)\pm},\notag\\
&\Gamma_{(k)}^{\pm}=\mp4\left(\sum_{i=1}^{k-1}\mathrm{Re}(\textrm{i}\gamma_{i})-\sum_{i=k+1}^{N}
\mathrm{Re}(\textrm{i}\gamma_{i})\right),|y_{k}^{(1)}\rangle=\left(
                       \begin{array}{c}
                         1 \\
                         \xi_{k} \\
                       \end{array}
                     \right),~~|y_{k}^{(2)}\rangle=\left(
                       \begin{array}{c}
                         \xi_{k} \\
                         1 \\
                       \end{array}
                     \right).
\end{aligned}
\end{equation}
Finally, in terms of the expression \eqref{AF-36}, we obtain the asymptotic behavior \eqref{AF-43-1}.
\end{proof}

According to Theorem 4.3, the interaction between different types of localized waves is elastic. In order to describe the above results more intuitively, we display the following dynamic behaviors figures by taking $a=1.0, b=0.5, A=\frac{5}{12}$.

\begin{figure}
\begin{center}
\includegraphics[width=5.8cm,height=4.0cm,angle=0]{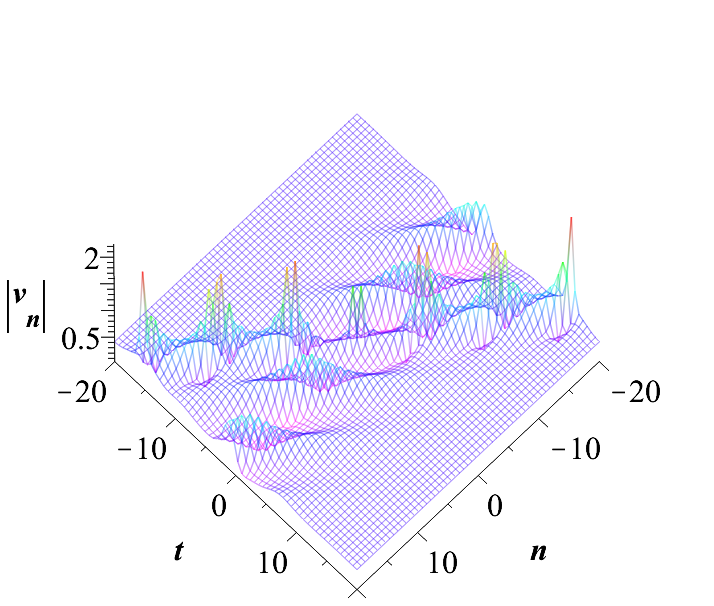}
~~\includegraphics[width=5.8cm,height=4.0cm,angle=0]{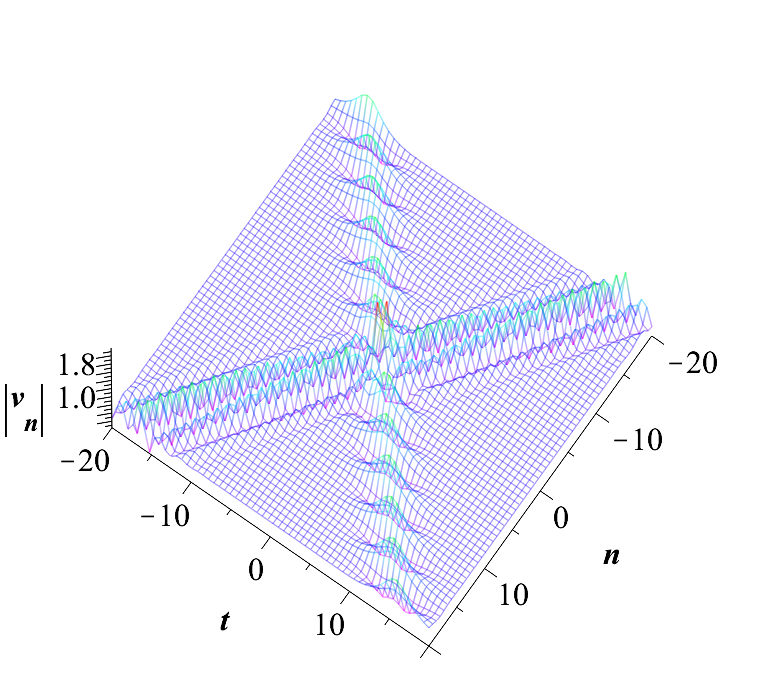}\\
$\hspace{0em}\textbf{(a)}
\hspace{18em}\textbf{(b)}$\\
\includegraphics[width=6.0cm,height=4.0cm,angle=0]{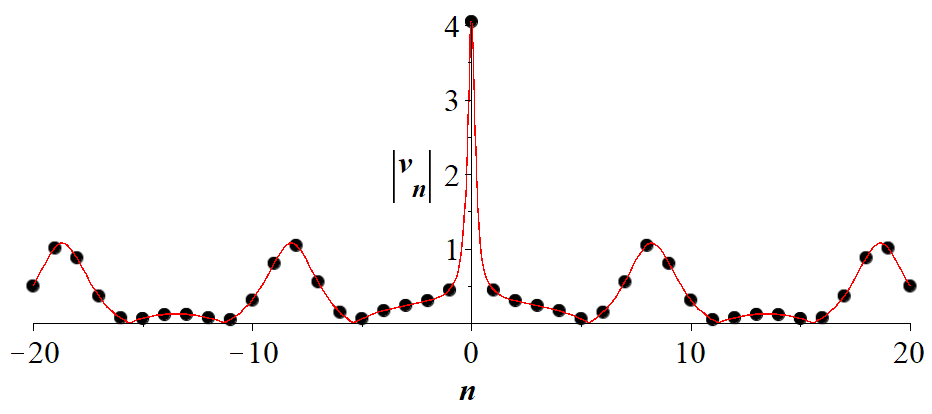}
~~\includegraphics[width=6.0cm,height=4.0cm,angle=0]{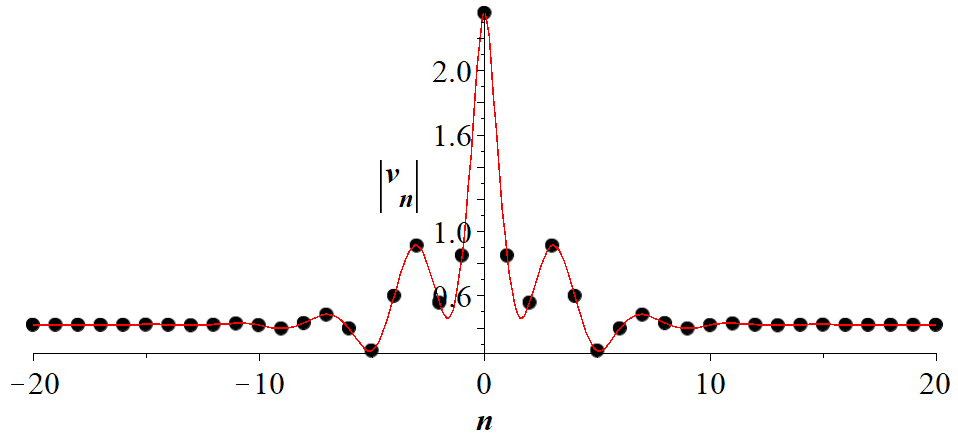}\\
$\hspace{0em}\textbf{(c)}
\hspace{18em}\textbf{(d)}$
  \end{center}
\end{figure}
{\small Figure 4\quad The interaction between two breather waves in \eqref{AF-43-1} with $B=0$:
$\textbf{(a)}$ $z_{1}=1.3, z_{2}=1.8$;
$\textbf{(b)}$ $z_{1}=1+0.9\textrm{i}, z_{2}=1-0.9\textrm{i}$;
wave propagation pattern of the wave along with the $n$ axis:
$\textbf{(c)}$ $\max|v_{n}^{[2]}|\approx4.05$;
$\textbf{(d)}$ $\max|v_{n}^{[2]}|\approx2.36$.}

\begin{figure}
\begin{center}
\includegraphics[width=5.8cm,height=4.0cm,angle=0]{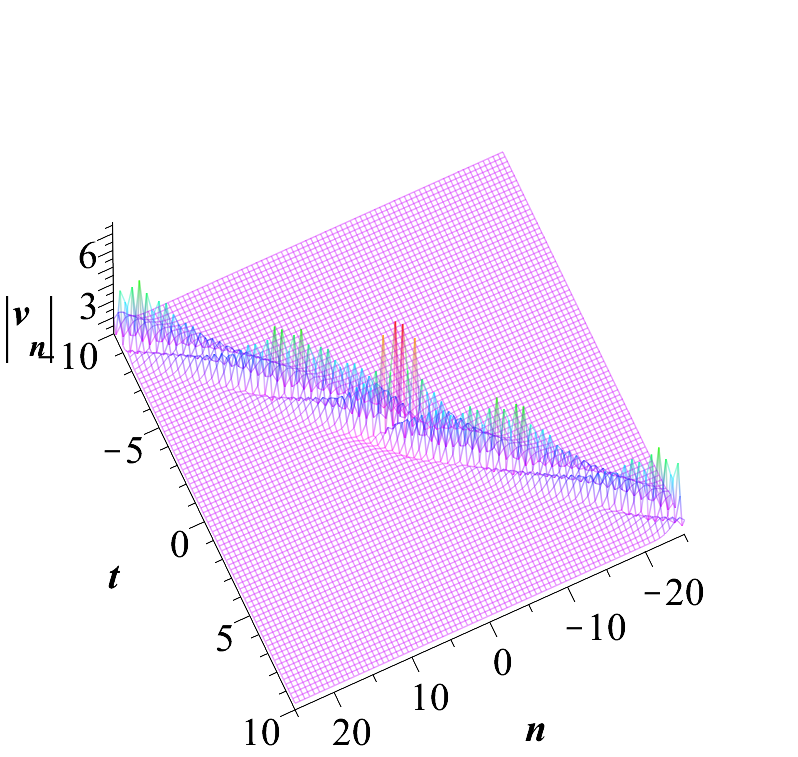}
~~\includegraphics[width=5.8cm,height=4.0cm,angle=0]{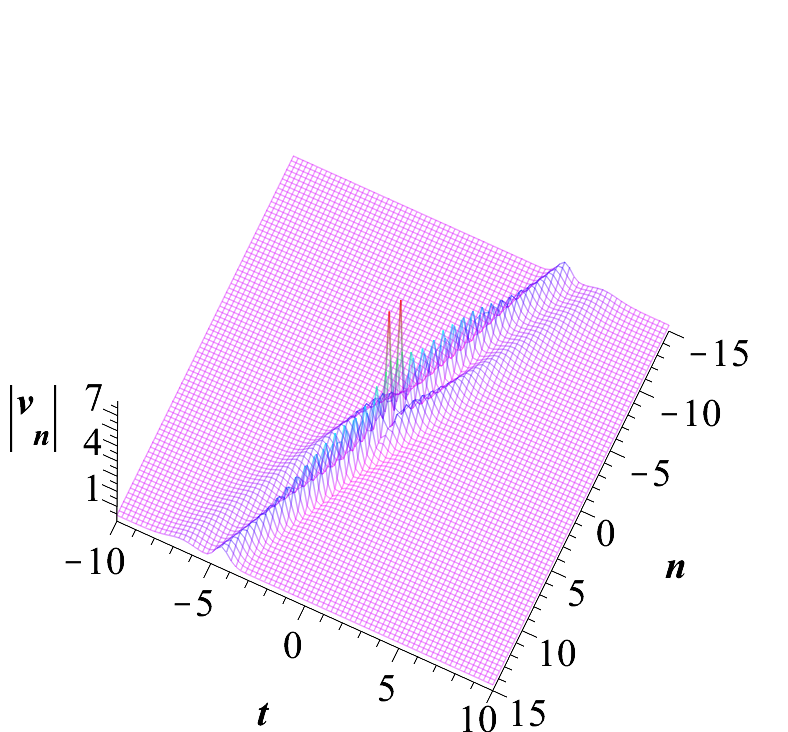}\\
$\hspace{0em}\textbf{(a)}
\hspace{18em}\textbf{(b)}$\\
\includegraphics[width=5.8cm,height=4.0cm,angle=0]{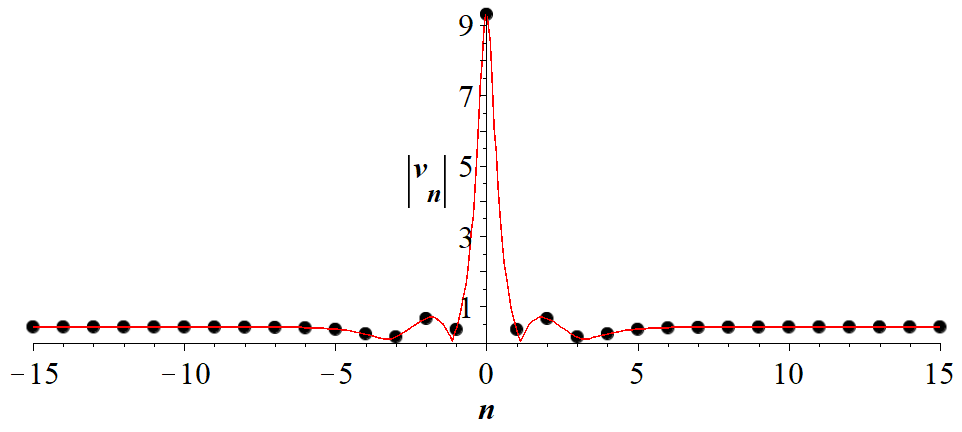}
~~\includegraphics[width=5.8cm,height=4.0cm,angle=0]{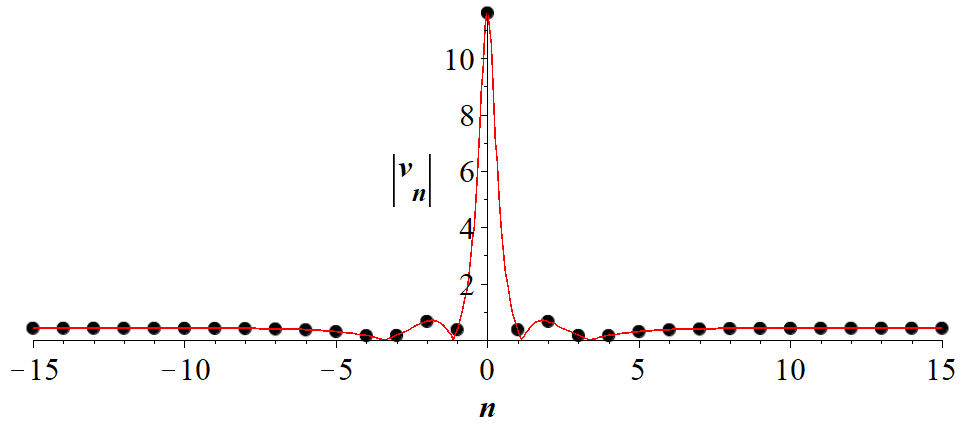}\\
$\hspace{0em}\textbf{(c)}
\hspace{18em}\textbf{(d)}$
  \end{center}
\end{figure}
{\small Figure 5\quad The interaction between a breather wave and a soliton as well as the interaction between two solitons in \eqref{AF-43-1} with $B=\frac{\pi}{2}$:
$\textbf{(a)}$ $z_{1}=\frac{7}{4}, z_{2}=\frac{7}{4}-\textrm{i}$;
$\textbf{(b)}$ $z_{1}=\frac{7}{4}, z_{2}=\frac{9}{4}$;
wave propagation pattern of the wave along with the $n$ axis:
$\textbf{(c)}$ $\max|v_{n}^{[2]}|\approx9.3$;
$\textbf{(d)}$ $\max|v_{n}^{[2]}|\approx11.6$.}

\begin{figure}
\begin{center}
\includegraphics[width=5.8cm,height=4.0cm,angle=0]{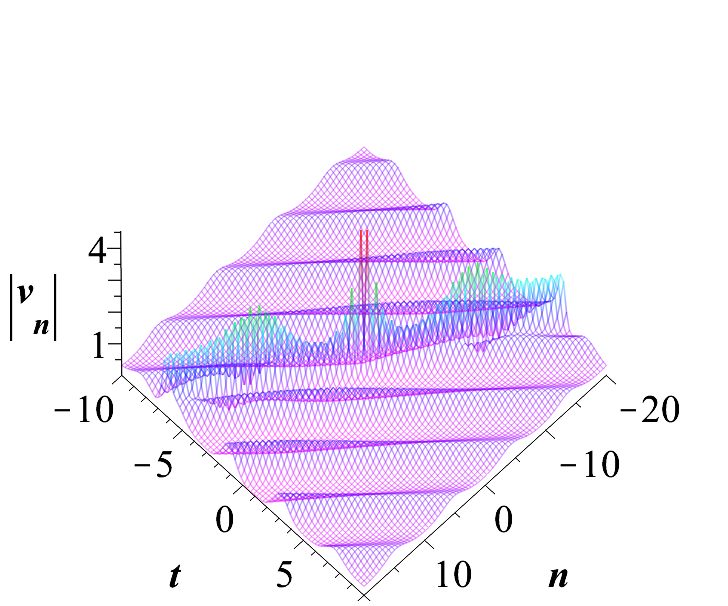}
~~\includegraphics[width=5.8cm,height=4.0cm,angle=0]{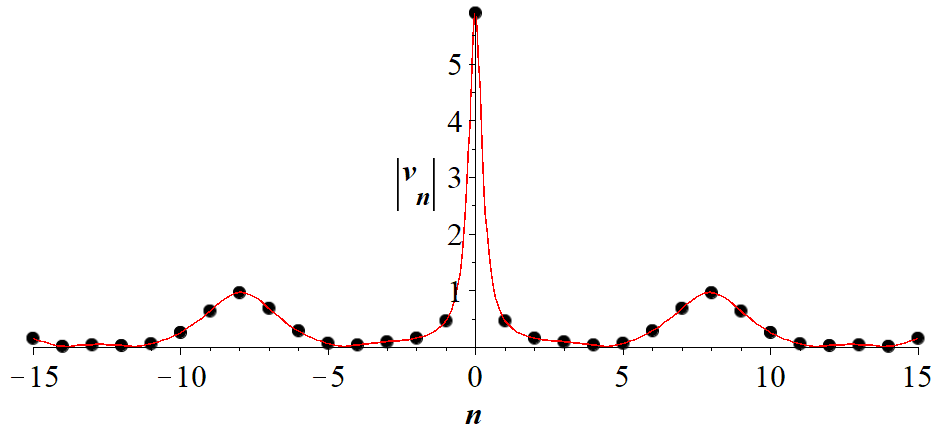}\\
$\hspace{0em}\textbf{(a)}
\hspace{18em}\textbf{(b)}$
  \end{center}
\end{figure}
{\small Figure 6\quad The interaction between a breather wave and a periodic wave in \eqref{AF-43-1} with $B=\frac{\pi}{2}, z_{1}=\frac{5}{4}, z_{2}=\frac{9}{4}$;
wave propagation pattern of the wave along with the $n$ axis: $\max|v_{n}^{[2]}|\approx5.89$.}
\vspace{0.2cm}

From the above figures, there are more abundant types of soliton solutions under the modulational stable background, compared to the modulational unstable one.
In addition, they are all elastic. Under the modulational unstable background, we obtain the interaction between breathers. Figs. 4(a) and (b) describe the interactions between Akhmediev breather and Kuznetsov-Ma breather as well as two Tajiri-Watanabe breathers, respectively.
Under the modulational stable background, the interactions between a Tajiri-Watanabe breather and a W-shape soliton as well as two W-shape solitons are depicted in Figs. 5(a) and (b).
We get the wave propagation pattern of the interactions between a Tajiri-Watanabe breather and a periodic solution from Fig. 6(b).

\subsection{High-order rogue waves}

The elementary matrix solution for the Lax pair \eqref{AF-4} is normalized as
%We normalize the elementary matrix solution of the Lax pair \eqref{AF-4} with the form
\begin{equation}\label{AF-44}
\begin{split}
\Phi(n,t,z)=&r^{n-n_{0}}\textrm{e}^{(a\sin B+b\cos B)A^{2}(t-t_{0})}G\textrm{e}^{\eta\sigma_{3}}G^{-1}\\
=&r^{n-n_{0}}\textrm{e}^{(a\sin B+b\cos B)A^{2}(t-t_{0})}\\
&\times\left(
                                 \begin{array}{cc}
                                   \cosh(\chi)-\frac{(1-z^{2})\sinh(\chi)}{2z\omega} & \frac{A\sinh(\chi)}{\omega} \\
                                   -\frac{A\sinh(\chi)}{\omega} & \cosh(\chi)+\frac{(1-z^{2})\sinh(\chi)}{2z\omega} \\
                                 \end{array}
                               \right),
\end{split}
\end{equation}
with
\begin{equation}\nonumber
\begin{aligned}
&\Phi(n_{0},t_{0},z)=\mathbb{I}_{2},\notag\\
&\chi=(n-n_{0})\ln\zeta+\delta\omega(t-t_{0})+\tilde{c}.
\end{aligned}
\end{equation}
When $z\in\mathbb{C}\setminus \{0,\infty\}$, $\Phi(n,t,z)$ is analytic. The branch points $z= r \pm A$ or $z=-r \pm A$ are its removable
singularities. For given $n$ and $t$, we have the vector solutions
\begin{align}\label{AF-45}
\left(
  \begin{array}{c}
    f(n,t,z) \\
    g(n,t,z) \\
  \end{array}
\right)=\Phi(n,t,z)\left(
                           \begin{array}{c}
                             1 \\
                             (r+A)z \\
                           \end{array}
                         \right).
\end{align}
We write these solutions in \eqref{AF-45} as
\begin{equation}\nonumber
\begin{aligned}
f(n,t,z)=\sum_{i=0}^{\infty}f_{1}^{[i]}(z-z_{1})^{i},~
g(n,t,z)=\sum_{i=0}^{\infty}g_{1}^{[i]}(z-z_{1})^{i},~
\tilde{c}=\sum_{i=0}^{\infty}c_{i}(z-z_{1})^{i},
\end{aligned}
\end{equation}
where
\begin{equation}\nonumber
 \begin{split}
z=z_{1}\equiv r+A,~~f_{1}^{[i]}=f_{1}^{[i]}(n,t),~~g_{1}^{[i]}=g_{1}^{[i]}(n,t).
 \end{split}
\end{equation}
According to Theorem 4.2 and Eq. \eqref{AF-45}, the single soliton solutions arrive at the maximum crest when $n=t=0$. In fact,
it is also suitable for $N$-soliton solutions and high-order rogue waves because the Darboux-BT can be iterated recursively.
\vspace{0.2cm}

\noindent
\textbf{Theorem 4.4}
Given $(n_{0},t_{0}) = (0,0)$ and $\tilde{c}= 0$, we obtain the maximum value $M_{i}$ for the elementary $i$th-order rogue waves \eqref{AF-46} with the following recursion expression
\begin{equation}\nonumber
\begin{aligned}
M_{1}&=\frac{(r+A)^{3}}{2}-\frac{(r-A)^{3}}{2},\notag\\
M_{i}&=\frac{(r+A)^{2}}{2}\left(\sqrt{1+M_{i-1}^{2}}+M_{i-1}\right)\notag\\
&-\frac{(r-A)^{2}}{2}\left(\sqrt{1+M_{i-1}^{2}}-M_{i-1}\right),
~~i\in\mathbb{Z},~i\geq2.
\end{aligned}
\end{equation}
\vspace{0.2cm}

Based on Propositions 3.2 and 3.3, we deduce the following theorem.
\vspace{0.2cm}

\noindent
\textbf{Theorem 4.5}
The high-order rogue wave solutions for the discrete Hirota equation \eqref{AF-3} are defined as the form of determinant
\begin{align}\label{AF-46}
v_{n}^{[N]}=A\frac{\det(H)}{\det(T)},
\end{align}
where
\begin{equation}\nonumber
\begin{aligned}
&T=K_{1}^{\dagger}\mu K_{1}+J_{2}^{\dagger}\mu J_{2},
~~H=K_{2}^{\dagger}\mu K_{2}+J_{1}^{\dagger}\mu J_{1}+F^{\dagger}K_{1,1}^{\dagger}J_{1,1},\notag\\
&\mu=\left(\frac{1}{(i-1)!(j-1)!}\frac{d^{i+j-2}}{dk^{i-1}dy^{j-1}}\left(\frac{1}{k^{2}y^{2}-1}
\right)|_{y=\bar{z}_{1},k=z_{1}}\right)_{1\leq i,j\leq N},\notag\\
&J_{1}=f_{1}^{[0]}\mathbb{I}_{N}+\sum_{i=1}^{N-1}f_{1}^{[j]}E^{j},~~J_{2}=z_{1}J_{1}+J_{1}E,~~E=(\delta_{i,j+1})_{1\leq i,j\leq N},\notag\\
&K_{1}=g_{1}^{[0]}\mathbb{I}_{N}+\sum_{i=1}^{N-1}g_{1}^{[j]}E^{j},~~K_{2}=z_{1}K_{1}+K_{1}E,\notag\\
&F=(\bar{z}_{1})^{-1}\mathbb{I}_{N}-(\bar{z}_{1})^{-2}E+(\bar{z}_{1})^{-3}E^{2}
+\cdots+(-1)^{N-1}(\bar{z}_{1})^{-N}E^{N-1},
\end{aligned}
\end{equation}
$K_{1,1}$ and $J_{1,1}$ denote the first row of matrices $K_{1}$ and $J_{1}$ respectively.
\vspace{0.2cm}

We next exhibit dynamic behaviors and wave propagation patterns of the rational solution by plugging $N=1, a=1.0, b=1.0, n_{0}=t_{0}=\tilde{c}=0, A=\frac{11}{60}$ into the formula \eqref{AF-46}.

\begin{figure}
\begin{center}
\includegraphics[width=5.8cm,height=4.0cm,angle=0]{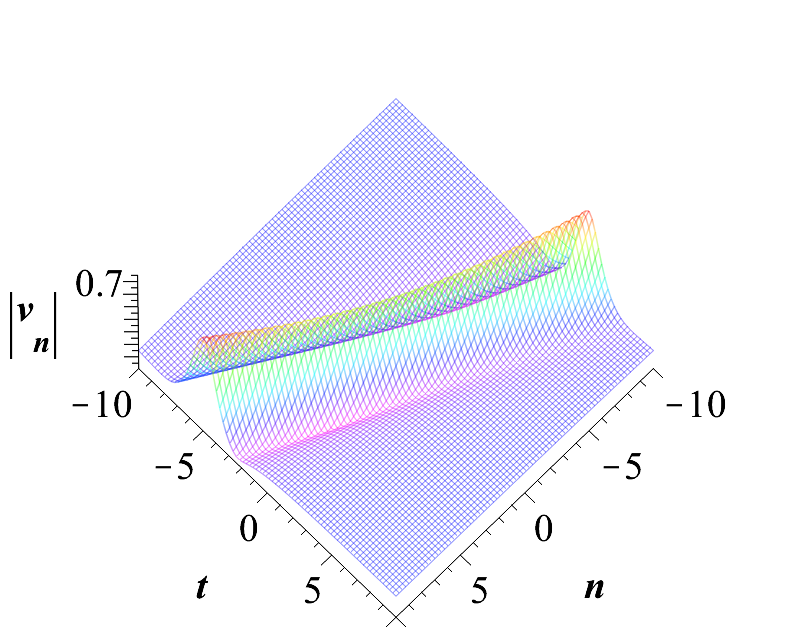}
~~\includegraphics[width=5.8cm,height=4.0cm,angle=0]{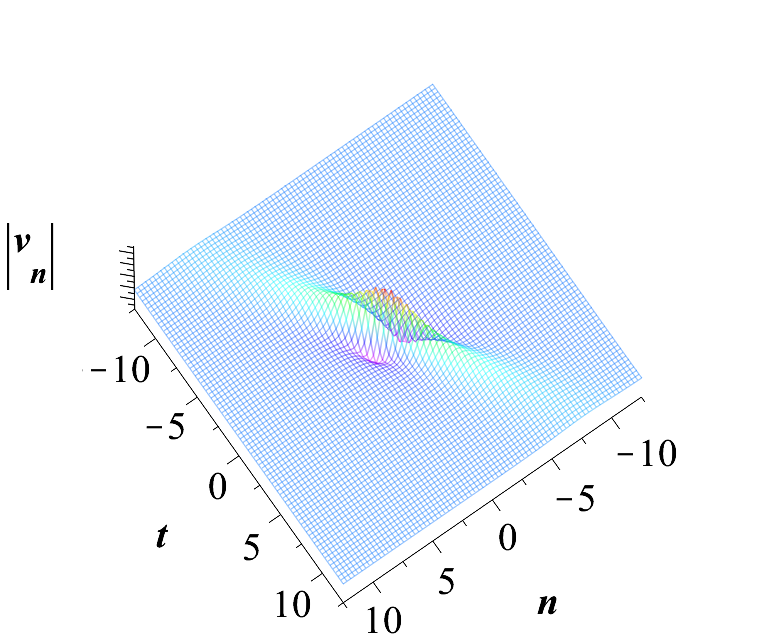}\\
$\hspace{0em}\textbf{(a)}
\hspace{18em}\textbf{(b)}$\\
\includegraphics[width=5.8cm,height=4.0cm,angle=0]{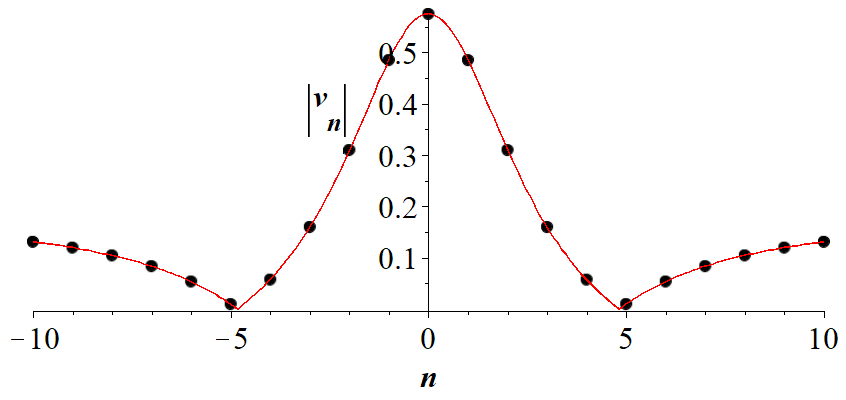}
~~\includegraphics[width=5.8cm,height=4.0cm,angle=0]{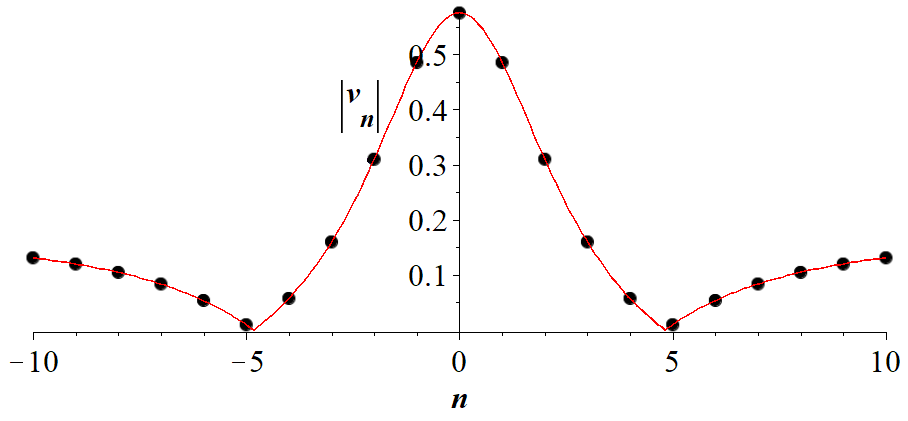}\\
$\hspace{0em}\textbf{(c)}
\hspace{18em}\textbf{(d)}$
  \end{center}
\end{figure}
{\small Figure 7\quad The rational solution \eqref{AF-46} with $N=1$:
$\textbf{(a)}$ $B=\frac{\pi}{2}$;
$\textbf{(b)}$ $B=0$;
wave propagation pattern of the wave along with the $n$ axis:
$\textbf{(c)}$ $\max|v_{n}^{[1]}|\approx0.57$;
$\textbf{(d)}$ $\max|v_{n}^{[1]}|\approx0.57$.}
\vspace{0.2cm}

From Figs. 7(a) and (b), the rational solution \eqref{AF-46} is a W-shape soliton solution in the context of modulational stable; otherwise, it is first-order rogue wave in the context of modulational unstable.
\vspace{0.2cm}

When $N > 1$, we take $a=1.0, b=0.3, B=0, n_{0}=t_{0}=0$ into the formula \eqref{AF-46} and draw the dynamic patterns of $N=3$ and $N=4$.

\begin{figure}
\begin{center}
\includegraphics[width=4.6cm,height=3.4cm,angle=0]{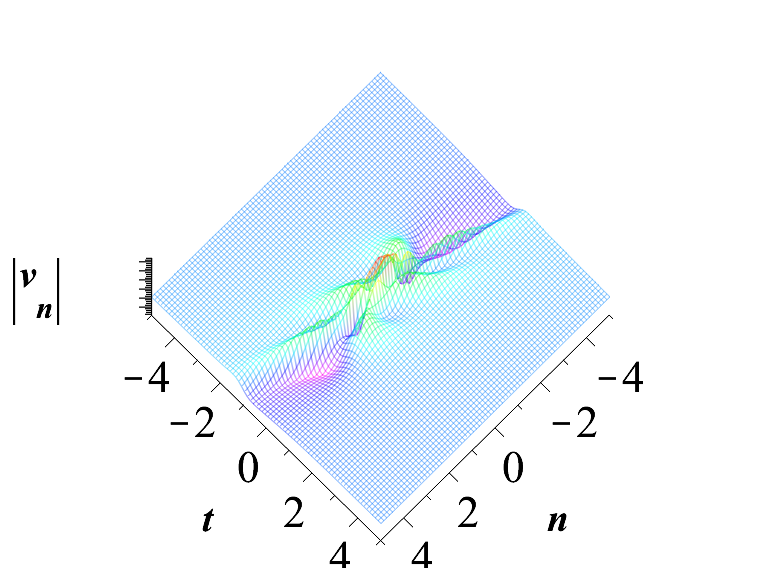}
~~\includegraphics[width=4.6cm,height=3.4cm,angle=0]{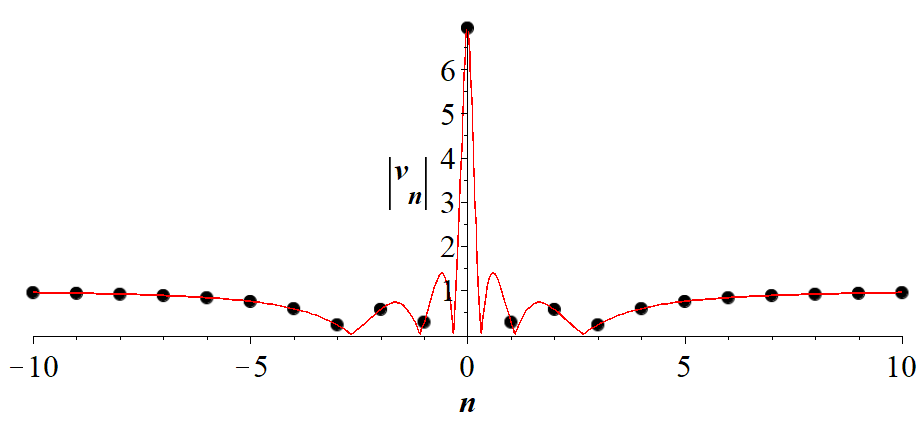}
~~\includegraphics[width=4.6cm,height=3.4cm,angle=0]{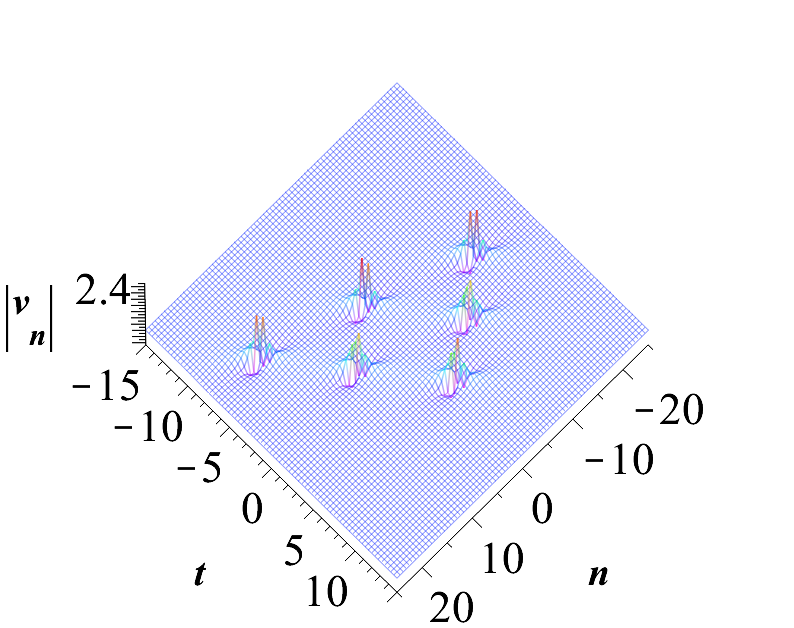}\\
$\hspace{0em}\textbf{(a)}
\hspace{14em}\textbf{(b)}
\hspace{14em}\textbf{(c)}$
\end{center}
\end{figure}
{\small Figure 8\quad 3rd-order rogue waves with $ A=\frac{23}{60}$:
$\textbf{(a)}$ $\tilde{c}=0$;
$\textbf{(b)}$ wave propagation pattern of the wave along with the $n$ axis, $\max|v_{n}^{[3]}|\approx6.84$;
$\textbf{(c)}$ the wave with triangle shape when $\tilde{c}=400\textrm{i}\epsilon \ln(\zeta_{1}(\epsilon))$.}

\begin{figure}
\begin{center}
\includegraphics[width=4.6cm,height=3.4cm,angle=0]{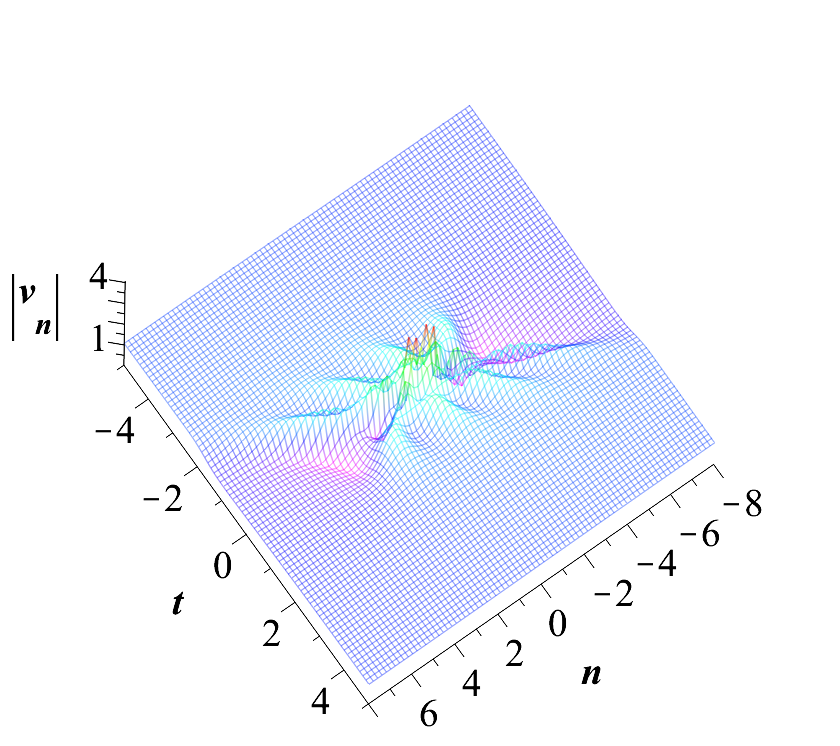}
~~\includegraphics[width=4.6cm,height=3.4cm,angle=0]{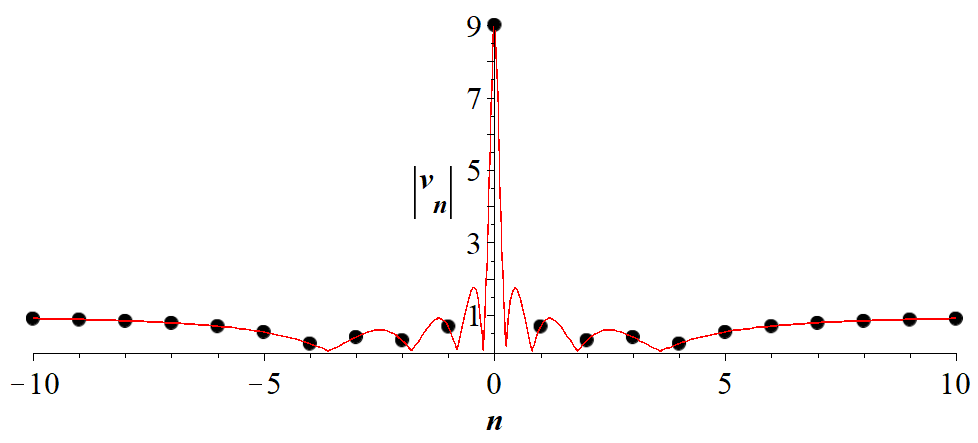}
~~\includegraphics[width=4.6cm,height=3.4cm,angle=0]{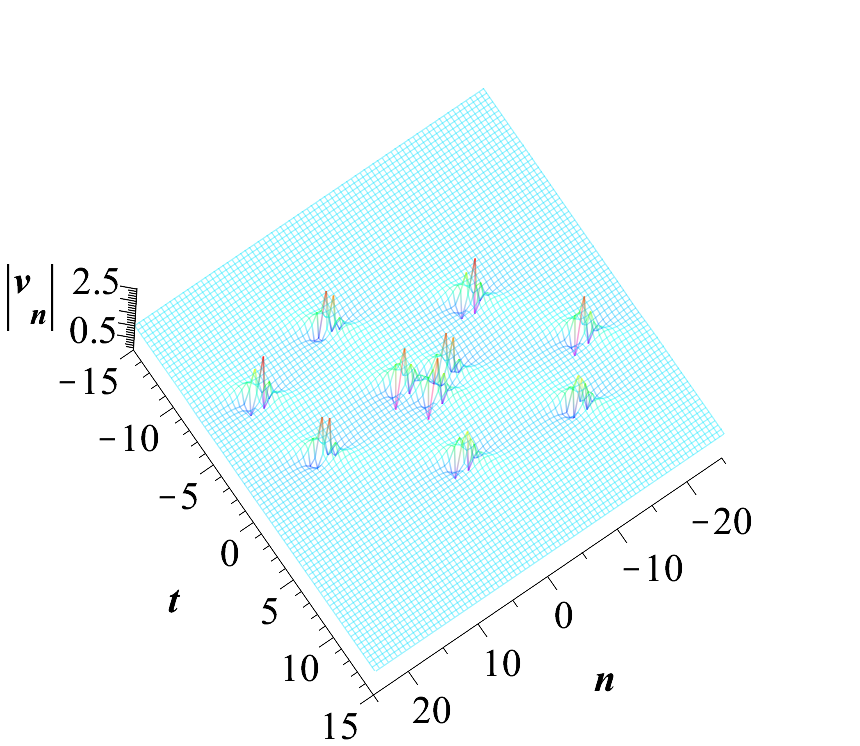}\\
$\hspace{0em}\textbf{(a)}
\hspace{14em}\textbf{(b)}
\hspace{14em}\textbf{(c)}$
  \end{center}
\end{figure}
{\small Figure 9\quad 4th-order rogue waves with $ A=\frac{18}{55}$:
$\textbf{(a)}$ $\tilde{c}=0$;
$\textbf{(b)}$ wave propagation pattern of the wave along with the $n$ axis, $\max|v_{n}^{[4]}|\approx9.02$;
$\textbf{(c)}$ the wave with ellipse shape when $\tilde{c}=10\textrm{i}\epsilon \ln(\zeta_{1}(\epsilon))$.}
\vspace{0.2cm}

Figs. 8(a) and 9(a) describe 3rd-order and 4th-order elementary rogue waves, respectively. Figs. 8(b) and 9(b) perfectly admit the formula of  maximum crest values in Theorem 4.4.
We can obtain dynamic images of high-order rogue waves with different shapes by changing the value of $\tilde{c}$, as shown in Figs. 8(c) and 9(c).

\section{Conclusions}
In this paper, we have investigated the discrete Hirota equation \eqref{sdHE} with the NZBCs \eqref{AF-1} via the robust IST. In contrast to classical IST, the formulas of these obtained rational solutions are more compact. In addition, these solutions as well as interactions between solitons and breathers are analyzed graphically.
Then we calculate the maximum amplitude for crest of these solutions.

Although we only consider Eq. \eqref{sdHE} as a particular example, the robust IST we present in this paper can be readily generalized to a wide range of continuous and discrete integrable systems, such as the discrete sine-Gordon equation, discrete Kundu-Eckhaus equation, the Ablowitz-Ladik equation etc. More importantly, there exists how to relate properties from the discrete model to continuous one in the corresponding continuous limit? The question is handled by considering continuous limit theory of integrable discrete model, where integrable properties computed from the discrete model will in general be approximations to their continuous counterparts in the corresponding continuous limit. This idea will be left for future discussions.\\

\noindent
\textbf{Acknowledgments:} This work of the first author was supported by the National Natural Science Foundation of China (No.12271129) and the China Scholarship Council (No.202206
120152). The work of the second author was supported by the National Natural Science Foundation of China (No.12201622). The work of the third author was
supported by the National Natural Science Foundation of China (No.12271129).

%\end{CJK*}
%\end{document}

% Converted from Microsoft Word to LaTeX
% by Chikrii SoftLab Word2TeX converter (version 2.4)
% Copyright (C) 1999-2001 Kirill A. Chikrii, Anna V. Chikrii
% Copyright (C) 1999-2001 Chikrii SoftLab.
% All rights reserved.
% http://www.word2tex.com/
% mailto: info@word2tex.com, support@word2tex.com

% Warning: You are using UNREGISTERED Chikrii SoftLab Word2TeX!
%          In UNREGISTERED mode some restrictions will apply.
%          For more information please visit http://www.word2tex.com/
% YOU CAN USE THIS FILE WITH THE SOLE PURPOSE OF EVALUATING Word2TeX.

%\documentclass [12pt]{article}

%\begin{document}

\end{document}